\documentclass[
nofootinbib,
nobibnotes,
bibnotes,
]{revtex4-2}  
\usepackage{subcaption}
\usepackage{graphicx}
\usepackage{dcolumn}
\usepackage{bm}
\usepackage{amsmath}
\usepackage{amssymb}
\usepackage{color}
\usepackage{ulem}
\usepackage[mathscr]{euscript}
\usepackage{upgreek}

\usepackage{physics}
\newcommand*{\ud}{\mathrm{\,d}}

\begin{document}

\preprint{APS/123-QED}

\title{System-bath correlations and finite-time operation enhance the efficiency of a dissipative quantum battery}

\author{Daniel Feliú}
 \email{feliuaraya.daniel@gmail.com}
\author{Felipe Barra}%
 \email{fbarra@dfi.uchile.cl}
\affiliation{Departamento de F\'{i}sica, Facultad de Ciencias F\'{i}sicas y Matem\'{a}ticas, Universidad de Chile, 837.0415 Santiago, Chile}

\date{\today}

\begin{abstract}
The reduced state of a small system strongly coupled to a thermal bath may be athermal and used as a small battery once disconnected. If the disconnecting process is too slow, the coupling between the battery and the bath weakens, and at some point, the battery will be in a thermal state that can not be used as a battery. Thus, the unitarily extractable energy (a.k.a ergotropy) decreases with the disconnection time. The work required to disconnect the battery also depends on the disconnection time. We study the efficiency of this battery, defined as the ratio between the ergotropy to the work cost of disconnecting and connecting the battery back to the bath to close the cycle, as a function of the disconnecting time in the Caldeira-Leggett model of a quantum battery. We consider two scenarios. In the first scenario, we assume that the discharged battery is uncorrelated to the bath at the connecting time and find that the efficiency peaks at an optimal disconnecting time. In the second scenario, the discharged battery is correlated to the bath, and find that the optimal efficiency corresponds to an instantaneous disconnection. On top of these results, we analyze various thermodynamic quantities for these Caldeira-Leggett quantum batteries that allow us to express the first and second laws of thermodynamics in the mentioned cycles in simple form despite the system-bath initial correlations and strong coupling regime of the working device. 
\end{abstract}

\maketitle


\section{\label{sec:intro}Introduction}

For a system to fulfill the role of a battery, regardless of its nature or the laws of physics that govern it, it must be able to store energy that can later be extracted. 
A good battery stores energy for a prolonged period, i.e., it has no leaks, and it is easy to extract the stored energy and charge it back. These challenges seem inherent to any battery, like the recently proposed isolated~\cite{alicki2013,binder2015,campaioli2017,ferraro2018,campaioli2023} or open~\cite{carrega2020,carrasco2022,farina2019,gherardini2020,mitchison2021,kamin2020,xu2021,xu2023,downing2023,campaioli2023,barra2019} quantum batteries. A particular type of open quantum battery was studied in ~\cite{hovhannisyan2020,barra2022}. They rely on a strong interaction with a bath and favorably occupy this dissipative interaction that provides stability and charge to the battery. 
However, its functioning is inherently dissipative. The agent that manipulates the battery by disconnecting and reconnecting it to the bath must expend energy that significantly surpasses the energy that a quantum system can extract from the battery. Although the utility of the battery depends on the fact that the energy that the agent uses to connect and disconnect is readily available to him, and the one that the battery provides to another quantum system is valuable, we would like to improve its efficiency, i.e., increase the ratio of extracted work to the agent energy cost. 
In this article, we address this problem following~\cite{hovhannisyan2020} and~\cite{barra2022}, based on two methodologies: intervening both in the coupling by considering a finite-time disconnecting protocol and in the correlations between the battery and the bath. For both, we obtain a favorable response.

Although optimizing the charging process of a battery has its own merit, the problem we address confronts essential questions in quantum thermodynamics~\cite{gemmer2009,vinjanampathy2016,binder2018,deffner2019}. In general, a unified theoretical framework for the laws of thermodynamics that applies to the quantum regime is lacking, as evidenced by the number of different proposals~\cite{strasberg2020,rivas2020,talkner2020,esposito2010,manzano2016,bera2017,strasberg2019,dolatkhah2020,landi2021,elouard2023}. The main difficulty is dealing with systems that are correlated and/or strongly coupled to their environment. Despite this, one can formulate a consistent thermodynamic description in specific processes and setups where correlations and coupling are present. For instance, an apparent violation of the second law, in which energy flows from a colder to a hotter system, is explained by initial correlations between the systems~\cite{vitagliano2018,micadei2019}. We also understand how correlations between quantum systems influence the work extracted and the efficiency of thermal engines and impact the charging of a quantum battery ~\cite{dillenschneider2009,niedenzu2018,altintas2015,pozas2018,binder2015,ferraro2018,andolina2019,campaioli2017,gyhm2022}. Similarly with the effect of strong coupling between systems on the power and efficiency of a machine, that can be favorably or unfavorably~\cite{gallego2014,gelbwaser2015,katz2016,perarnau2018,shirai2021,latune2023}. This work is also a contribution towards the understanding of thermodynamics in the regime of correlated and strongly coupled systems. We study cycles for a dissipative quantum battery based on the Caldeira-Leggett model in which the cyclic condition of the working fluid implies an equality that allows us to formulate the first and second laws of thermodynamics of the cycle in a standard way~\cite{fermi2012,callen1998}.

The rest of the article is organized as follows. In section~\ref{sec:cycles}, we introduce the cycle for the battery. Two different scenarios are presented. In the first, the bath is a bipartite system, and the recharging process occurs with the battery in contact with a fresh thermal bath (similarly to the case studied in ~\cite{hovhannisyan2020}). In the second, the battery is reconnected to the same bath from which it was disconnected (similarly to the case studied in ~\cite{barra2022}). In~\cite{hovhannisyan2020,barra2022}, the battery is instantaneously disconnected from that bath. Here, we introduce a finite-time disconnecting protocol. Then, in section~\ref{sec:level2}, the Caldeira-Leggett model and its properties are discussed. In section~\ref{sec:level2p}, we apply the scenarios discussed in section~\ref{sec:cycles}, to the Caldeira-Leggett model. We present the techniques used for the numerical implementation and the numerical results in section~\ref{sec:level1}. We summarize and conclude in section~\ref{sec:conclu}. Two appendices complement our article. In the first appendix, we show numerically that the battery fulfills the cyclic condition in the second scenario. In the second appendix, we derive an equality for the cycle in the Caldeira-Leggett model, which is crucial in formulating the thermodynamic laws.

\section{\label{sec:cycles}Charge-Discharge cycles in dissipative quantum batteries}

The energy coupling is not negligible for a small system in equilibrium with a bath. So, its reduced state, the so-called mean-force Gibbs state, is not necessarily passive~\cite{pusz1978,lenard1978}, and it may be possible to extract work from it. This observation is the starting point for the dissipative quantum batteries explored in~\cite{hovhannisyan2020} and~\cite{barra2022}. Three main advantages follow from that. The first is that the battery is passively charged by thermalization. The second is that the stability of the equilibrium state ensures that the charge is preserved without leaks in the storing energy phase. The third is that one does not need precise time control as in~\cite{ferraro2018} to disconnect from its charger, the bath. We disconnect the battery from the charger to extract the energy. This step is necessary since a cyclic unitary process can extract no energy from a system in a Gibbs state~\cite{pusz1978,lenard1978}. Considering that the energy is extracted with a unitary process, we characterize its charge with the ergotropy, $\mathscr{W}$, introduced in~\cite{allahverdyan2004}.

In a nutshell, these batteries work in the following cycle. They start initially coupled to a bath; the entire system bath is in the Gibbs state, but the reduced state of the battery is the mean-force Gibbs state. We disconnect the battery from the bath to extract the ergotropy. Then, we reconnect it to the bath and wait until the thermalization process charges the battery again. 
Because the coupling is strong, disconnecting and connecting the battery to the bath has a work cost of $W_{\rm d}$ and $W_{\rm c}$ to the agent that manipulated the battery. The efficiency of the device is the ratio between the ergotropy and the sum of these works, $\eta:=\mathscr{W}/(W_{\rm d}+W_{\rm c}).$

The following two subsections will detail two possible scenarios to implement the battery cycle as mentioned above. In the first, the bath is a bipartite system; thus, the total system is tripartite. In the second scenario, the total system is bipartite. This allows us to analyze the role of system bath correlations in thermodynamic quantities, particularly the device's efficiency.

For the sake of simplicity of the expressions used and of the numerical implementation, which we will discuss later, we are considering units in which $\hbar=k_{\rm B}=1$, where $\hbar$ is the reduced Planck constant and $k_{\rm B}$ is the Boltzmann constant.

\subsection{Tripartite composite system}
\label{sec:Tcs}

The four-stroke cycle for the battery follows from the total system unitary dynamics ruled by a Hamiltonian of the form
\[
\hat{\mathcal{H}}_{\rm tot}(t)=\hat{H}_{\rm S}+\hat{H}_{\rm B}+\hat{H}_{\rm B'}+\hat{V}_{\rm SB}(t)+\hat{V}_{\rm SB'}(t),
\] 
where $\hat{H}_{\rm S}$ is the battery Hamiltonian, $\hat{H}_{\rm B}+\hat{H}_{\rm B'}$ the bath Hamiltonian, which is bipartite and each subsystem $\rm B$ or $\rm B'$ can couple to the battery with the interaction term $\hat{V}_{\rm SB}(t)$ or $\hat{V}_{\rm SB'}(t)$ but not among themselves. 
We consider both parts to be identical, i.e., $\hat{H}_{\rm B}=\hat{H}_{\rm B'}$. Also, the time-independent couplings between bath and system 
$\hat{V}_{\rm SB}^{0}=\hat{V}_{\rm SB'}^{0}.$

The cycle starts with the battery interacting with the first bath, $\rm B$. Thus at the beginning of the process, $t=0$,  $\hat{V}_{\rm SB'}(0)=0$, $\hat{V}_{\rm SB}(0)=\hat{V}_{\rm SB}^{0}$ and 
\begin{equation}
\hat{\mathcal{H}}_{\rm tot}(0)=\hat{H}_{\rm S}+\hat{H}_{\rm B}+\hat{H}_{\rm B'}+\hat{V}_{\rm SB}^{0}\equiv\hat{H}_{\rm SB}+\hat{H}_{\rm B'}.
\label{Htri}
\end{equation}
The initial state of the tripartite system is the thermal state with inverse temperature $\beta$,
\begin{equation}
\hat{\uptau}=\frac{e^{-\beta \hat{\mathcal{H}}_{\rm tot}(0)}}{\mathcal{Z}}=\hat{\tau}_{\rm SB}\otimes\hat{\tau}_{\rm B'}.
\label{tautri}
\end{equation}
The reduced states $\hat{\tau}_{\rm SB}=e^{-\beta\hat{H}_{\rm SB}}/Z_{\rm SB}$ of the battery-$\rm B$ system and  $\hat{\tau}_{\rm B'}=e^{-\beta\hat{H}_{\rm B'}}/Z_{\rm B'}$ of $\rm B'$ are also thermal. 
Importantly, the battery and B in the state $\hat{\tau}_{\rm SB}$ are correlated, and tracing over bath $\rm B$ gives the battery's reduced mean-force Gibbs state $\hat{\tau}_{\rm MF}=\rm Tr_{\rm B}\hat{\tau}_{\rm SB}$. All these states are normalized and $\mathcal{Z}=Z_{\rm SB}Z_{\rm B'}$, $Z_{\rm SB}$ and $Z_{\rm B'}$ are their respective partition functions.
We disconnect the battery from the bath, by monotonically decreasing a time-dependent parameter in $\hat{V}_{\rm SB}(t)$ until its value $\hat{V}_{\rm SB}(t_{\rm d})=0$. 
This first stroke has a time duration $t_{\rm d}$, and during this lapse of time, $\hat{V}_{\rm SB'}(t)=0$, so the bath $\rm B'$ has a passive role. Moreover, $\hat{\tau}_{\rm B'}$ is invariant under its free evolution. The disconnecting work is 
\begin{equation}
W_{\rm d}(t_{\rm d})={\rm Tr}[(\hat{H}_{\rm S}+\hat{H}_{\rm B})\hat{\rho}_{\rm SB}(t_{\rm d})]-{\rm Tr}[(\hat{H}_{\rm S}+\hat{H}_{\rm B}+\hat{V}_{\rm SB}^0)\hat{\tau}_{\rm SB}],
\label{Wdtri}
\end{equation}
where 
\begin{equation}
\hat{\rho}_{\rm SB}(t_{\rm d})=\hat{U}_{\rm d}(t_{\rm d})\hat{\tau}_{\rm SB}\hat{U}_{\rm d}(t_{\rm d})^{\dagger}
\label{rhoSBtd}
\end{equation}
and $\hat{U}_{\rm d}(t_{\rm d})$ the evolution operator determined by $\hat{H}_{\rm S}+\hat{H}_{\rm B}+\hat{V}_{\rm SB}(t)$. The utility of the battery in the cycles will be determined by the fact its state $\hat{\rho}_{\rm S}(t_{\rm d})={\rm Tr}_{\rm B}\hat{\rho}_{\rm SB}(t_{\rm d})$ has some ergotropy at the end of the disconnection process. Once disconnected, the bath $\rm B$ plays no role in this process, so we trace it out. The second stroke is the extraction process in which a local unitary $\hat{U}_{\mathscr{W}}$ extracts the ergotropy, $\mathscr{W}(t_{\rm d})$, of the battery. For simplicity (mainly notational), we assume this process to be instantaneous. With the battery in its passive state 
\begin{equation}
\hat{\rho}_{\rm S}^{\rm p}(t_{\rm d})=\hat{U}_{\mathscr{W}}\hat{\rho}_{\rm S}(t_{\rm d})\hat{U}_{\mathscr{W}}^\dagger,
\label{UW}
\end{equation}
we perform the third stroke in which we connect the battery to the bath $\rm B'$. To simplify the analysis, we consider an instantaneous connection process in which $\hat{V}_{\rm SB'}(t)$ changes instantaneously from $\hat{V}_{\rm SB'}(t_{\rm d})=0$ to $\hat{V}_{\rm SB'}(t_{\rm d})=\hat{V}_{\rm SB'}^0$, while the state $\hat{\rho}_{\rm S}^{\rm p}(t_{\rm d})\otimes \hat{\tau}_{\rm B'}$ remains unchanged. Thus, the connecting work is
\begin{equation*}
W_{\rm c}(t_{\rm d})={\rm Tr}[\hat{V}_{\rm SB'}^0\hat{\rho}_{\rm S}^{\rm p}(t_{\rm d})\otimes \hat{\tau}_{\rm B'}]. 
\end{equation*}
We assume that after this quench-like protocol~\cite{Rigol2009,perarnau2018,hovhannisyan2020}, the autonomous unitary evolution of the composed system 
\begin{equation}
\hat{U}_{\rm ch}(t)\hat{\rho}_{\rm S}^{\rm p}(t_{\rm d})\otimes \hat{\tau}_{\rm B'}\hat{U}_{\rm ch}^\dagger(t)
\label{Uchbi}
\end{equation}
with $\hat{U}_{\rm ch}(t)=\exp\left(-i[\hat{H}_{\rm S}+\hat{H}_{\rm B'}+\hat{V}_{\rm SB'}^0]t\right)$ brings the battery to the equilibrium state. That is,
\[
\lim_{t\rightarrow+\infty}{\rm Tr}_{\rm B'}[\hat{U}_{\rm ch}(t)\hat{\rho}_{\rm S}^{\rm p}(t_{\rm d})\otimes \hat{\tau}_{\rm B'}\hat{U}_{\rm ch}^\dagger(t)]=\hat{\tau}_{\rm MF},
\]
which coincides with the initial state because $\hat{H}_{\rm B}=\hat{H}_{\rm B'}$ and 
$\hat{V}_{\rm SB}^{0}=\hat{V}_{\rm SB'}^{0}$.

\subsubsection{Energetic and efficiency of the cycle}

The state of the tripartite system after ergotropy extraction and also after the connection, which we do in a quench at time $t=t_{\rm d}$, is
\begin{equation}
    \hat{\Upomega}(t_{\rm d})=\left[\hat{U}_{\mathscr{W}}\otimes\hat{\mathbb{I}}_{\rm B}\hat{\rho}_{\rm SB}(t_{\rm d})(\hat{U}_{\mathscr{W}}\otimes\hat{\mathbb{I}}_{\rm B})^{\dagger}\right]\otimes\hat{\tau}_{\rm B'}
    \label{tristateendext}
\end{equation}
(where $\hat{\mathbb{I}}_{\rm B}$ is the identity operator on the bath Hilbert space). The state for $t>t_{\rm d}$ is 
\begin{equation}
    \hat{\Upomega}(t_{\rm d};t)=
    \hat{\mathcal{U}}_{\rm ch}(t) \hat{\Upomega}(t_{\rm d})\hat{\mathcal{U}}_{\rm ch}^{\dagger}(t),
    \label{Omegatri}
\end{equation}
which is determined by the autonomous evolution operator
\begin{equation}
\hat{\mathcal{U}}_{\rm ch}(t)=\exp(-i\hat{H}_{\rm B}t)\otimes\hat{U}_{\rm ch}(t) =\exp(-i\hat{H}_{\rm B}t)\otimes\exp(-i\hat{H}_{\rm SB'}t).
\label{Utrich}
\end{equation}
The initial Hamiltonian 
\begin{equation}\label{Hin}
\hat{\mathcal{H}}=\hat{H}_{\rm S}+\hat{H}_{\rm B}+\hat{H}_{\rm B'}+\hat{V}_{\rm SB}^{0}=-\beta^{-1}\ln\hat{\uptau}-\beta^{-1}\ln \mathcal{Z}
\end{equation}
and the final Hamiltonian 
\begin{equation}\label{Hfin}
\hat{\mathcal{H}}'=\hat{H}_{\rm S}+\hat{H}_{\rm B}+\hat{H}_{\rm B'}+\hat{V}_{\rm SB'}^{0}=-\beta^{-1}\ln\hat{\uptau}'-\beta^{-1}\ln \mathcal{Z}'
\end{equation}
can be written in terms of the initial state $\hat{\uptau}$ and of the auxiliary state $\hat{\uptau}'=\hat{\tau}_{\rm B}\otimes\hat{\tau}_{\rm SB'}$ introduced in the second equality for $\hat{\mathcal{H}}'$, where we also have $\mathcal{Z}'=Z_{\rm SB'}Z_{\rm B}$. 

Because energy is conserved by the autonomous evolution in the charging process, the energy change 
\begin{equation}
    \begin{split}
\Delta \mathcal{E}^{\rm cyc}(t_{\rm d};t)
    &:={\rm Tr}[\hat{\mathcal{H}}'\hat{\Upomega}(t_{\rm d};t)]-{\rm Tr}[\hat{\mathcal{H}}\hat{\uptau}]
    \end{split}
\label{deltaEtri}
\end{equation}
of the tripartite system from $t=0$, where the state was $\hat{\uptau}$
to a time $t>t_{\rm d}$, equals the dissipated work defined as $W_{\rm diss}(t_{\rm d}):=W_{\rm d}(t_{\rm d})+W_{\rm c}(t_{\rm d})-\mathscr{W}(t_{\rm d})$. Then, substituting the expressions \eqref{Hin} and \eqref{Hfin} for $\hat{\mathcal{H}}$ and $\hat{\mathcal{H}}'$ into equation \eqref{deltaEtri}, we obtain
\begin{equation}
    \begin{split}
    W_{\rm diss}(t_{\rm d})&=\Delta \mathcal{E}^{\rm cyc}(t_{\rm d};t)\\
    &=-\beta^{-1}{\rm Tr}[\hat{\Upomega}(t_{\rm d};t)\ln\hat{\uptau}']+\beta^{-1}{\rm Tr}[\hat{\uptau}\ln\hat{\uptau}]\\ &=\beta^{-1}D\left(\hat{\Upomega}(t_{\rm d};t)||\hat{\uptau}'\right)-\beta^{-1}S(\hat{\uptau})+\beta^{-1}S(\hat{\Upomega}(t_{\rm d};t))\\
    &=\beta^{-1}D\left(\hat{\Upomega}(t_{\rm d};t)||\hat{\uptau}'\right)\geq0.
    \end{split}
    \label{43}
\end{equation}
Here, $S(\hat{\rho}):=-{\rm Tr}[\hat{\rho}\ln\hat{\rho}]$ denotes the von Neumann entropy of a state $\hat{\rho}$, and $D(\hat{\rho}||\hat{\sigma}):=\rm Tr[\hat{\rho}\ln\hat{\rho}]-\rm Tr[\hat{\rho}\ln\hat{\sigma}]\geq0$ the quantum relative entropy, with equality holding when $\hat{\rho}=\hat{\sigma}$. In the final equality in \eqref{43}, we have utilized the invariance of the von Neumann entropy under unitary transformations, which implies $S(\hat{\Upomega}(t_{\rm d};t))=S(\hat{\uptau})$. Additionally, we have employed $\mathcal{Z}=\mathcal{Z}'$, stemming from the equalities $\hat{H}_{\rm B}=\hat{H}_{\rm B'}$ and $\hat{V}_{\rm SB}^{0}=\hat{V}_{\rm SB'}^{0}.$ 
From the inequality $W_{\rm diss}(t_{\rm d})\geq0$ it follows that the efficiency of the total process fulfills
\begin{equation} 
\eta(t_{\rm d})=\frac{\mathscr{W}(t_{\rm d})}{W_{\rm d}(t_{\rm d})+W_{\rm c}(t_{\rm d})}\leq 1\;\forall t_{\rm d}. 
\label{etatri}
\end{equation}

\subsection{Bipartite composed system}
\label{Bcs}

The second case considered differs from the previous in that, after extracting the ergotropy of the battery, we connect it to the same bath $\rm B$. Thus, the second bath $\rm B'$ can be neglected. In this bipartite system, the disconnecting work and the ergotropy have the same values as in the tripartite case, but the connecting work differs. 
In fact, as discussed in \cite{barra2022}, there is a family of unitaries that extract the ergotropy. This implies that if the continuous variable $\theta$ parametrize  the family, i.e.,  $\hat{U}_{\mathscr{W}}(\theta)$ then 
$\hat{U}_{\mathscr{W}}(\theta)\hat{\rho}_{\rm S}(t_{\rm d})\hat{U}_{\mathscr{W}}^\dagger(\theta)=\hat{\rho}_{\rm S}^{\rm p}(t_{\rm d})$ for all $\theta$. Thus, this parameter is irrelevant for the tripartite scenario. However, in the present case, 
the battery and the bath may be correlated (they were before the disconnection process, and though these correlations change, they are not likely to disappear). The state of the composite system after the ergotropy is extracted from the battery is 
\begin{equation}
\hat{\Omega}(t_{\rm d},\theta)=(\hat{U}_{\mathscr{W}}(\theta)\otimes \hat{\mathbb{I}}_{\rm B})\hat{\rho}_{\rm SB}(t_{\rm d})(\hat{U}_{\mathscr{W}}(\theta)\otimes \hat{\mathbb{I}}_{\rm B})^\dagger,
\label{statebitheta}
\end{equation}
and the connecting work (again, we consider a quench) is
\[
W_{\rm c}(t_{\rm d},\theta)={\rm Tr}[V_{\rm SB}^{0}\hat{\Omega}(t_{\rm d},\theta)],
\]
which depends on $\theta$, implying that the efficiency also depends on $\theta$,
\begin{equation}
    \eta(t_{\rm d},\theta)=\frac{\mathscr{W}(t_{\rm d})}{W_{\rm d}(t_{\rm d})+W_{\rm c}(t_{\rm d},\theta)}.
    \label{etabi}
\end{equation}
Note that the $\theta$ dependence disappears if $\hat{\rho}_{\rm SB}(t_{\rm d})$ were a product state. We assume that the unitary dynamics that follow after this connecting process thermalizes the battery to its state  $\hat{\tau}_{\rm MF}$.

\subsubsection{Energetic and efficiency of the cycle}

Similar to the tripartite scenario, the dissipated work of the process in the bipartite case, $W_{\rm diss}(t_{\rm d},\theta)$, equals the energy change, $\Delta E^{\rm cyc}(t_{\rm d},\theta;t)$, of the bipartite system in the process from $t=0$ to $t>t_{\rm d}$, i.e., 
\begin{widetext}
\begin{equation}
\begin{split}
    W_{\rm diss}(t_{\rm d},\theta)&:=W_{\rm d}(t_{\rm d})+W_{\rm c}(t_{\rm d},\theta)-\mathscr{W}(t_{\rm d})\\
    &={\rm Tr}[(\hat{H}_{\rm S}+\hat{V}_{\rm SB}^{0}+\hat{H}_{\rm B})(\hat{\Omega}(t_{\rm d},\theta;t)-\hat{\tau}_{\rm SB})]\\
    &:=\Delta E^{\rm cyc}(t_{\rm d},\theta;t),
    \end{split}
    \label{WdissBI}
\end{equation}
\end{widetext}
where 
\begin{equation*}
    \hat{\Omega}(t_{\rm d},\theta;t)=U_{\rm ch}(t)\hat{\Omega}(t_{\rm d},\theta)U^\dagger_{\rm ch}(t), 
\label{statebit}
\end{equation*}
and $\hat{\Omega}(t_{\rm d},\theta)$ given in Eq.\eqref{statebitheta}. Also, we have
\begin{equation}
    \begin{split}
    \Delta E^{\rm cyc}(t_{\rm d},\theta;t)&=\beta^{-1}D(\hat{\Omega}(t_{\rm d},\theta;t)||\hat{\tau}_{\rm SB})\\
    &\geq0.
    \end{split}
    \label{geq2}
\end{equation}
This last inequality was expected because, in this scenario, is a consequence of the passivity of the global initial Gibbs state~\cite{pusz1978,lenard1978}. In this 
sense, the equivalent inequality for the tripartite case is less trivial. 

As in the tripartite case, this inequality implies that the efficiency in Eq. \eqref{etabi} satisfies
\begin{equation*}
\eta(t_{\rm d},\theta)\leq1\;\forall (t_{\rm d},\theta).
\end{equation*}

Similar scenarios were considered in previous works~\cite{hovhannisyan2020,barra2022}, but only with instantaneous disconnecting processes. 
Here, we explore the role of the disconnection time $t_{\rm d}$ in the thermodynamics of the battery. If $t_{\rm d}=0$, i.e., the disconnection is also a quench, the tripartite cycle corresponds to the one studied in~\cite{hovhannisyan2020}
and the bipartite cycle to the one studied in~\cite{barra2022}.

We study these two scenarios with a Caldeira-Leggett model for the system-bath. For this model, it is known that the charging dynamics in the first scenario indeed bring the battery to its mean-force state. For the second scenario, there is no proof of that result, but we present strong numerical evidence that it happens in the thermodynamic limit. Thus, all the assumptions made above are met when working with this model. 
Moreover, we find a remarkable property that allows us to make some non-trivial remarks about the thermodynamics of the cycle.  

\section{\label{sec:level2}Caldeira-Leggett model and mean-force (MF) Gibbs state}

We aim to understand how switching on and off the interaction between the battery and environment (or bath) affects the thermodynamics and the battery's performance~\cite{barra2019,hovhannisyan2020,barra2022}. In particular, the role of the time $t_{\rm d}$ it takes to switch it off.  
To achieve this goal, we modeled the battery by a quantum oscillator (referred to as the \textit{central oscillator}) that linearly interacts with a bath consisting of a collection of uncoupled harmonic oscillators with a time-dependent coupling strength. In the next section (section~\ref{sec:level2p}), we introduce this time dependence in the Caldeira-Leggett model.
Here, we review the relevant results of the usual time-independent Caldeira-Leggett model that rules the charging dynamics and the statistical properties of the battery before disconnecting it from the bath. 

The Caldeira-Leggett total Hamiltonian~\cite{caldeira1983} of the composed system (battery + bath + interaction) is given by
\begin{equation}
    \hat{H}_{\rm SB}=\underbrace{\left(\frac{\hat{P}_{0}^{2}}{2m_0}+\frac{m_0\omega_0^2}{2}\hat{Q}_{0}^{2}\right)}_{\hat{H}_{\rm S}}+\underbrace{\sum_{k=1}^{N}\left(\frac{\hat{P}_{k}^{2}}{2m_k}+\frac{m_k\omega_k^2}{2}\hat{Q}_{k}^{2}\right)}_{\hat{H}_{\rm B}}+\underbrace{\frac{m_0\omega_{\rm R}^{2}}{2}\hat{Q}_{0}^{2}-\hat{Q}_{0}\sum_{k=1}^{N}g_{k}\hat{Q}_{k}}_{\hat{V}_{\rm SB}^0},
\label{Hcl}
\end{equation}
where the first parenthesis is the Hamiltonian of the central oscillator ($\hat{H}_{\rm S}$), the first sum up to $N$ is the bath Hamiltonian ($\hat{H}_{\rm B}$), and the last two terms --which we will collectively call $\hat{V}^{0}_{\rm SB}$-- are the interaction Hamiltonian (with negative sign) and a \textit{renormalization term} or \textit{counter-term} with the \textit{renormalization frequency}, $\omega_{\rm R}$, defined as
\begin{equation}
    \omega^2_{\rm R}:=\sum_{k=1}^{N}\frac
{g_{k}^{2}}{m_0 m_k \omega_{k}^{2}}.
\label{omegaR}
\end{equation}
The counter-term is introduced to prevent the central oscillator from suffering a frequency shift due to its interaction with the bath \cite{weiss2012,dittrich1998,breuer2002}. This ensures that the unique reduced steady state of the central oscillator, in the high-temperature limit, is the thermal Gibbs state corresponding to the natural oscillator's frequency, $\propto e^{-\beta\hat{H}_{\rm S}}$, which interacts with an initially thermal bath with inverse temperature $\beta:=1/T$ in a purely dissipative dynamics \cite{correa2017,weiss2012}. 

Consider that the bath is initially in the thermal state $\hat{\tau}_{\rm B}=e^{-\beta\hat{H}_{\rm B}}/Z_{\rm B}$ and uncorrelated with the central oscillator, i.e., the total initial state is $\hat{\rho}_{\rm tot}(0)=\hat{\rho}_{\rm S}(0)\otimes\hat{\tau}_{\rm B}$ (as we do in the tripartite scenario). Then, in the thermodynamic limit ($N\rightarrow+\infty$), the dynamics of the central oscillator can be described by a \textit{generalized Langevin equation} \cite{weiss2012,dittrich1998,breuer2002}
\begin{equation}
    \dv[2]{\hat{Q}_{0}(t)}{t}+\int_{0}^{t}\ud t'\gamma(t-t')\dv{\hat{Q}_{0}(t')}{t'}\\+\omega_{0}^{2}\hat{Q}_{0}(t)=\frac{\hat{\xi}(t)}{m_0}-\gamma(t)\hat{Q}_{0}(0),
\label{Langevin}
\end{equation}
where 
\begin{equation}
\gamma(t):=\begin{cases}
    \frac{1}{m_{0}}\sum_{k}\frac{g_{k}^{2}}{m_{k}\omega_{k}^{2}}\cos(\omega_{k}t),& \text{if } t>0\\
    0,& \text{if } t<0
\end{cases}
\label{gamma}
\end{equation}
is the memory-friction kernel and 
\begin{equation}
\hat{\xi}(t):=\sum_{k}g_{k}\left[\hat{Q}_{k}\cos(\omega_{k}t)+\frac{\hat{P}_{k}}{m_{k}\omega_{k}}\sin(\omega_{k}t)\right]
\label{noise}
\end{equation}
is the noise operator. The memory kernel $\gamma(t)$ and the noise operator $\hat{\xi}(t)$ are related via the fluctuation-dissipation theorem.  One example of this relationship is given by the following expression: 
\begin{equation}
    \langle\hat{\xi}(t)\hat{\xi}(t')\rangle_{\hat{\tau}_{\rm B}}=\frac{m_0}{\pi}\int_{0}^{\infty}\ud \omega \mathfrak{Re}[\tilde{\gamma}(\omega)]\omega \left(\coth\left(\frac{\beta\omega}{2}\right)
    \cos[\omega(t-t')]-i\sin[\omega(t-t')]\right),
\label{eq10}
\end{equation}
where $\mathfrak{Re}[\tilde{\gamma}(\omega)]$ is the real part of the Fourier transform of the memory kernel\footnote{See the Fourier transform definition used in \cite{weiss2012}.}, which is related to \textit{the spectral density}
\begin{equation*}
J(\omega):=\frac{\pi}{2}\sum_{k=1}^{N}\frac
{g_{k}^{2}}{m_{k}\omega_{k}}\delta(\omega-\omega_{k})
\end{equation*}
by 
\begin{equation*}
    \mathfrak{Re}[\tilde{\gamma}(\omega)]=\frac{J(\omega)}{m_0\omega}.
\end{equation*}
The spectral density $J(\omega)$ encodes the behavior of the bath (\textit{Ohmic}, \textit{sub-Ohmic}, and \textit{super-Ohmic}) and the coupling with the central oscillator \cite{weiss2012}. Considering Eq. \eqref{eq10}, it is possible to calculate from Eq. \eqref{Langevin} using the Laplace transform method (see, for example, Ref. \cite{bialas2018}) that the stationary solutions ($t\rightarrow+\infty$) of the expectation values $\langle\hat{Q}_{0}^{2}(+\infty)\rangle_{\hat{\rho}_{\rm tot}(0)}:={\rm Tr}[\hat{Q}_{0}^{2}(+\infty)\hat{\rho}_{\rm S}(0)\otimes\hat{\tau}_{\rm B}]$ and $\langle\hat{P}_{0}^{2}(+\infty)\rangle_{\hat{\rho}_{\rm tot}(0)}:={\rm Tr}[\hat{P}_{0}^{2}(+\infty)\hat{\rho}_{\rm S}(0)\otimes\hat{\tau}_{\rm B}]$ are
\begin{equation}
        \langle\hat{Q}_{0}^{2}(+\infty)\rangle_{\hat{\rho}_{\rm tot}(0)}=\frac{1}{\pi m_0}\int_{0}^{\infty}\ud \omega\frac{\omega\mathfrak{Re}[\tilde{\gamma}(\omega)]}{|\alpha(\omega)|^{2}}\coth\left(\frac{\beta\omega}{2}\right)
        \label{Q2stationary}
\end{equation}
and 
\begin{equation}
        \langle\hat{P}_{0}^{2}(+\infty)\rangle_{\hat{\rho}_{\rm tot}(0)}=\frac{m_0}{\pi}\int_{0}^{\infty}\ud \omega\frac{\omega^{3}\mathfrak{Re}[\tilde{\gamma}(\omega)]}{|\alpha(\omega)|^{2}}\coth\left(\frac{\beta\omega}{2}\right),
        \label{P2stationary}
\end{equation}
with $\alpha(\omega):=\omega_{0}^{2}-\omega^{2}-i\omega\tilde{\gamma}(\omega)$. In addition, it is obtained that the expectation values 
\begin{equation}
\begin{split}
\langle\{\hat{Q}_{0}(+\infty),\hat{P}_{0}(+\infty)\}\rangle_{\hat{\rho}_{\rm tot}(0)}&=\langle\{\hat{P}_{0}(+\infty),\hat{Q}_{0}(+\infty)\}\rangle_{\hat{\rho}_{\rm tot}(0)}\\
&=0.
\end{split}
\label{offdiagonal}
\end{equation}
The expressions \eqref{Q2stationary}, \eqref{P2stationary}, and \eqref{offdiagonal} fully characterize the stationary state of the central oscillator. Moreover, $\langle\hat{Q}_{0}^{2}(+\infty)\rangle_{\hat{\rho}_{\rm tot}(0)}=\langle\hat{Q}_{0}^{2}\rangle_{\hat{\tau}_{\rm MF}}$, $\langle\hat{P}_{0}^{2}(+\infty)\rangle_{\hat{\rho}_{\rm tot}(0)}=\langle\hat{P}_{0}^{2}\rangle_{\hat{\tau}_{\rm MF}}$ and $\langle\{\hat{Q}_{0},\hat{P}_{0}\}\rangle_{\hat{\tau}_{\rm MF}}=\langle\{\hat{P}_{0},\hat{Q}_{0}\}\rangle_{\hat{\tau}_{\rm MF}}=0$ with~\cite{grabert1984},
\begin{equation}
    \hat{\tau}_{\rm MF}:=\rm Tr_{\rm B}\hat{\tau}_{\rm SB},
    \label{tauMF}
\end{equation}
where $\hat{\tau}_{\rm SB}=e^{-\beta\hat{H}_{\rm SB}}/Z_{\rm SB}$; with $Z_{\rm SB}={\rm Tr}[e^{-\beta\hat{H}_{\rm SB}}]$ the partition function of the composed system. Therefore, given the quadratic nature of the Caldeira-Leggett Hamiltonian \eqref{Hcl} and the Gaussian nature of $\hat{\tau}_{\rm MF}$ this steady state is equivalent to tracing out the bath degrees of freedom from the thermal equilibrium state with temperature $T$ of the composite system~\cite{adesso2014}. However, it is important to note that the global state of the composite system at long times does not correspond to $\hat{\tau}_{\rm SB}$. The state $\hat{\tau}_{\rm MF}$ in Eq. \eqref{tauMF} is known as the \textit{mean-force Gibbs state}. This state is a Gaussian state~\cite{serafini2017} that, equivalently, can be written in a Gibbsian exponential form with respect to the inverse of the initial temperature of the bath, $\beta$, and an effective temperature-dependent Hamiltonian (\textit{mean-force Hamiltonian}), $\hat{H}_{\rm MF}$, i.e., $\hat{\tau}_{\rm MF}\propto e^{-\beta\hat{H}_{\rm MF}}$. For a detailed description of the mean-force Gibbs state, see references \cite{talkner2020, trushechkin2022}.

The solutions \eqref{Q2stationary} and \eqref{P2stationary} depend on the election of a particular form for $J(\omega)$ to describe the continuous spectra of the bath and the coupling of this with the central oscillator. In this work, we choose an Ohmic spectral density, $J(\omega)\propto\omega$ if $\omega\rightarrow0$, with a Lorentz-Drude regularization for the numerical treatment of the discrete and large bath \cite{weiss2012}. This spectral density is given by
\begin{equation}
    J(\omega)=\frac{2m_{0}\gamma\omega}{1+(\omega/\omega_{\rm D})^{2}},
    \label{Lorentz-drude}
\end{equation}
which means that the Fourier transform of the memory kernel is of the form
\begin{equation*}
    \tilde{\gamma}(\omega)=\frac{2\gamma}{1-i\omega/\omega_{\rm D}},
    \label{Fouriergamma}
\end{equation*}
with $\omega_{\rm D}$ a cut-off frequency and $\gamma$ (without time dependency) a constant with frequency units. The cut-off frequency $\omega_{\rm D}$ ensures that $J(\omega)$ decays with increasing $\omega$. Therefore, the central oscillator cannot couple with bath oscillators with frequencies much higher than $\omega_{\rm D}$. Additionally, from a numerical perspective, if the considered bath is discrete and finite but sufficiently dense in frequencies to emulate the thermodynamic limit, then each constant $g_{k}$ in the Hamiltonian \eqref{Hcl} can be approximated in terms of the spectral density \eqref{Lorentz-drude} by means of the expression  
\begin{equation} 
g_{k}\approx\sqrt{\frac{4\gamma m_{0} m_{k}\omega_{k}^{2}\Delta_{k}}{\pi[1+(\omega_{k}/\omega_{\rm D})^{2}]}}, 
\label{couplings}
\end{equation}
where $\Delta_{1}:=\omega_{1}$ and $\Delta_{k\neq1}=\omega_{k}-\omega_{k-1}$ for a sample of frequencies defined by 
\begin{equation}
    \omega_{k}=a_{0}\tanh\left[\frac{\pi}{2}\frac{k}{N+1}\right],
    \label{frecuencies}
\end{equation}
with $a_{0}$ an adjustable parameter and $N$ the number of oscillators in the bath \cite{pucci2013}.

We will exploit those properties of the autonomous evolution of the compound system, and particularly the properties of the mean-force Gibbs state, to implement different cycles for the battery.

\section{\label{sec:level2p}Caldeira-Leggett Discharge-charge cycles}
We consider a battery cycle based on the Caldeira-Leggett model's unitary dynamics. The key in those global processes is that the disconnection process between the battery and the corresponding bath is mediated via a \textit{disconnection protocol}, $\lambda(t)$,
of duration $t_{\rm d}$ that enters in the interaction term as follows\footnote{Consider that the following transformation occurs in Eqs. \eqref{Hcl} and \eqref{omegaR}: $g_{k}\rightarrow\lambda(t)g_{k}$}:
\begin{equation}
   \hat{V}_{\rm SB}(t)= \lambda^{2}(t)\frac{m_0\omega_{\rm R}^{2}}{2}\hat{Q}_{0}^{2}-\lambda(t)\hat{Q}_{0}\sum_{k=1}^{N}g_{k}\hat{Q}_{k},
   \label{VSBt}
\end{equation}
where $\lambda(t)$ modulates the purely quadratic interaction, and a $\lambda^{2}(t)$ multiplies the renormalization term avoiding the instantaneous frequency shift. Eq. \eqref{VSBt} replaces $\hat{V}_{\rm SB}^0$ in Eq. \eqref{Hcl}. Note that for $\lambda(t)=1$, $\hat{V}_{\rm SB}(t)=\hat{V}_{\rm SB}^0$.

The disconnection protocol starts at $t=0$ with $\lambda(0)=1$, i.e., $\hat{V}_{\rm SB}(0)=\hat{V}_{\rm SB}^0$, and decreases monotonically until $t_{\rm d}$, the end of the process, where $\lambda(t_{\rm d})=0$; consequently, the interaction is turned off, i.e., $\hat{V}_{\rm SB}(t_{\rm d})=0$. Then, the state of the battery and the bath coupled to it evolve from $t=0$ to $t_{\rm d}$ with the unitary operator
\begin{equation}
    \hat{U}_{\rm d}(t_{\rm d})=\hat{\rm T}\exp\left(-i\int_{0}^{t_{\rm d}}\ud t' \hat{H}_{\rm SB}(t')\right)
    \label{Utrid}
\end{equation}
with $\hat{H}_{\rm SB}(t)$ the Caldeira-Leggett Hamiltonian in Eq. \eqref{Hcl} with the interaction $\hat{V}_{\rm SB}^0$ replaced by $\hat{V}_{\rm SB}(t)$ in Eq. \eqref{VSBt}.

The state of the battery disconnected from the bath is $\hat{\rho}_{\rm SB}(t_{\rm d})=\hat{U}_{\rm d}(t_{\rm d})\hat{\tau}_{\rm SB}\hat{U}_{\rm d}(t_{\rm d})^{\dagger}$, which is typically a correlated state
and the state of the battery is $\hat{\rho}_{\rm S}(t_{\rm d})={\rm Tr}_{\rm B}\hat{\rho}_{\rm SB}(t_{\rm d})$, which we expect to have some ergotropy at the end of the disconnection process.

In the following subsections, we consider some properties of the Caldeira-Leggett model that permit the simplification of some thermodynamic quantities, like the connecting work and its $\theta$ dependence in the bipartite scenario. We also discuss the thermodynamic implications of a non-trivial property involving the interaction energy and the initial and final states that follow from the condition of a cycle for the battery (convergence to the mean-force Gibbs state in the charging process).

Then, in the next section, we show how the quantities of interest, like the ergotropy of $\hat{\rho}_{\rm S}(t_{\rm d})$, or the connecting and disconneting works are computed for the Caldeira-Leggett battery. 

\subsection{Caldeira-Leggett tripartite composed system}

As discussed in section \ref{sec:Tcs}, in the tripartite scenario, the battery is connected to a copy of the bath $\rm B'$ in its Gibbs thermal state. The correlations between the first bath $\rm B$ and the battery do not play a role in this scenario. 
Because the interaction between the battery and the new bath is turned on abruptly, the connecting work is   
\begin{equation}
\begin{split}
    W_{\rm c}(t_{\rm d})
    &={\rm Tr}_{\rm S,B'}\left[\hat{V}_{\rm SB'}^{0}\hat{\rho}_{\rm S}^{\rm p}(t_{\rm d})\otimes\hat{\tau}_{\rm B'}\right]\\
    &=\frac{m_{0}\omega_{\rm R}^{2}}{2}{\rm Tr}_{\rm S}[\hat{Q}_{0}^{2}\hat{\rho}_{\rm S}^{\rm p}(t_{\rm d})]-{\rm Tr}_{\rm S}[\hat{Q}_{0}\hat{\rho}_{\rm S}^{\rm p}(t_{\rm d})]\sum_{k}g_{k}{\rm Tr}_{\rm B'}[\hat{Q}_{k}\hat{\tau}_{\rm B'}]\\
    &=\frac{m_{0}\omega_{\rm R}^{2}}{2}{\rm Tr}_{\rm S}[\hat{Q}_{0}^{2}\hat{\rho}_{\rm S}^{\rm p}(t_{\rm d})],
\end{split}
\label{Wctri}
\end{equation}
where $\hat{V}_{\rm SB'}^{0}=\hat{V}_{\rm SB}^{0}$ in Eq. \eqref{Hcl}.
In the last equality of \eqref{Wctri} we considered that ${\rm Tr}_{\rm B'}[\hat{Q}_{k}\hat{\tau}_{\rm B'}]=0$ $\forall k$. Also $\rm Tr_{\rm S}[\hat{Q}_{0}\hat{\rho}_{\rm S}^{\rm p}(t_{\rm d})]=0$, because $\hat{\rho}_{\rm S}^{\rm p}(t_{\rm d})$ is diagonal in the eigenbasis of $\hat{H}_{\rm S}$. 

For the battery to complete a cycle, the following condition must be met: $\lim_{t\rightarrow+\infty}{\rm Tr}_{\rm B,B'}[\hat{\Upomega}(t_{\rm d};t)]={\rm Tr}_{\rm B,B'}[\hat{\uptau}]$ [see Eqs. \eqref{tautri} and \eqref{Omegatri}]. This condition is guaranteed in the thermodynamic \cite{grabert1984,subasi2012}. Indeed, during the charging process, the battery solely interacts with the bath $\rm B'$, which begins in its thermal state, consequently, the reduced dynamics of the battery are accurately described by the Langevin equation \eqref{Langevin} in the thermodynamics limit, predicting that for any $\hat{\rho}_{\rm S}^{\rm p}(t_{\rm d})$,
    \begin{equation*}
        \begin{split}
            \hat{\tau}_{\rm MF}
            &=\lim_{t\rightarrow+\infty}\Tr_{\rm B'}[\hat{U}_{\rm ch}(t)\hat{\rho}_{\rm S}^{\rm p}(t_{\rm d})\otimes\hat{\tau}_{\rm B'}\hat{U}_{\rm ch}^{\dagger}(t)],
        \end{split}
    \end{equation*}
as we mentioned in section~\ref{sec:level2}. 
Therefore, the battery performs a cycle in the Caldeira-Leggett tripartite scenario.

\subsection{Caldeira-Leggett bipartite composed systems}

As discussed in section \ref{Bcs}, this scenario differs from the tripartite case in that the entire process (four strokes) occurs with the same bath $\rm B$, and the correlations between the battery and the bath affect the connecting work, charging process, and efficiency. Thus, the disconnecting work and the extracted ergotropy are the same than in the tripartite case. However, the arbitrary choice of the ergotropy-extracting operator now has a relevant role. 

For the Caldeira-Leggett battery, we can generate all these operators by composing $\hat{U}_{\mathscr{W}}$ used in the tripartite case with 
\begin{equation}
    \hat{U}_{\hat{n}_{0}}(\theta):=\exp(i\hat{n}_{0}\theta).
\label{Utheta}
\end{equation}
Here, $\hat{n}_{0}:=\hat{a}_{0}^{\dagger}\hat{a}_{0}$ is the number operator, where $a_{0}^{\dagger}=\sqrt{\frac{m_0\omega_0}{2}}\left(\hat{Q}_{0}-\frac{i}{m_0\omega_0}\hat{P}_{0}\right)$ and $\hat{a}_{0}=\sqrt{\frac{m_0\omega_0}{2}}\left(\hat{Q}_{0}+\frac{i}{m_0\omega_0}\hat{P}_{0}\right)$ are the ladder operators ($[\hat{a}_{0},\hat{a}_{0}^{\dagger}]=\hat{\mathbb{I}}_{\rm S}$) on the battery Hilbert space.
Therefore, the total extraction transformation is $\hat{U}_{\mathscr{W}}(\theta)=\hat{U}_{\hat{n}_{0}}(\theta)\hat{U}_{\mathscr{W}}$. Because $\hat{U}_{\hat{n}_{0}}(\theta)$ satisfies the commutation relation $[\hat{H}_{\rm S},\hat{U}_{\hat{n}_{0}}(\theta)]=0$, the unitary operator $\hat{U}_{\mathscr{W}}(\theta)$ extracts the ergotropy for any $\theta$. The state of the composed system after the extraction is $\hat{\Omega}(t_{\rm d},\theta)$ in Eq. \eqref{statebitheta}, which could have different correlations than $\hat{\rho}_{\rm SB}(t_{\rm d})$ due to the action of the operator $\hat{U}_{\mathscr{W}}(\theta)\otimes\hat{\mathbb{I}}_{\rm B}$. The reduced bath state remains unchanged after the extraction process. This can be directly observed by applying the Schmidt decomposition for pure bipartite states and acknowledging that any state can be decomposed using spectral decomposition (if mixed) \cite{nielsenBook,breuer2002}. Consequently, for any $t_{\rm d}$, the state $\hat{\rho}_{\rm SB}(t_{\rm d})$ in Eq. \eqref{rhoSBtd} can be decomposed as $\hat{\rho}_{\rm SB}(t_{\rm d})=\sum_{ij}c_{ij}\ket{\phi_i}\bra{\phi_j}\otimes\ket{\Phi_i}\bra{\Phi_j}$, where each $\ket{\phi_i}$ is a state of the system and each $\ket{\Phi_i}$ is a state of the bath. Then, by operating $\hat{U}_{\mathscr{W}}(\theta)\otimes\hat{\mathbb{I}}_{\rm B}$ and tracing over the battery, we have 
\begin{equation}
\begin{split}
{\rm Tr}_{\rm S}\hat{\Omega}_{\rm SB}(t_{\rm d},\theta)&={\rm Tr}_{\rm S}[\hat{U}_{\mathscr{W}}(\theta)\otimes\hat{\mathbb{I}}_{\rm B}\sum_{ij}c_{ij}\ket{\phi_i}\bra{\phi_j}\otimes\ket{\Phi_i}\bra{\Phi_j}\hat{U}_{\mathscr{W}}^{\dagger}(\theta)\otimes\hat{\mathbb{I}}_{\rm B}]\\
&=\sum_{ij}c_{ij}{\rm Tr}_{\rm S}[\hat{U}_{\mathscr{W}}(\theta)\ket{\phi_i}\bra{\phi_j}\hat{U}_{\mathscr{W}}^{\dagger}(\theta)]\otimes\ket{\Phi_i}\bra{\Phi_j}\\
&={\rm Tr}_{\rm S}\hat{\rho}_{\rm SB}(t_{\rm d}).
\end{split}
\label{ec42}
\end{equation}
Hence, the reduced state ${\rm Tr}_{\rm S}\hat{\Omega}_{\rm SB}(t_{\rm d},\theta)$ is independent of $\theta$.

After extracting the ergotropy, we connect the battery to the bath in a sudden quench process. The state of the composite system is conserved. We emphasize again that the state of the composite system has a phase dependence $(\theta)$, which will play an important role in calculating the work required to implement the process because $[\hat{V}_{\rm SB}^{0},\hat{U}_{\hat{n}_{0}}(\theta)]\neq0$. This poses a substantial difference with the tripartite cycles because here, we used $\theta$ as an optimization parameter. For the Caldeira-Leggett battery, we get
\begin{equation}
    \begin{split}
        W_{\rm c}(t_{\rm d},\theta)
        &={\rm Tr}[\hat{V}_{\rm SB}^{0}\hat{\Omega}(t_{\rm d},\theta)]\\
        &=\frac{m_{0}\omega_{\rm R}^{2}}{2}{\rm Tr}_{\rm S}[\hat{Q}_{0}^{2}\hat{\rho}_{\rm S}^{\rm p}(t_{\rm d})]-{\rm Tr}[\hat{Q}_{0}\sum_{k}g_{k}\hat{Q}_{k}\hat{\Omega}(t_{\rm d},\theta)].
    \label{Wcbi}
    \end{split}
\end{equation}
From Eq. \eqref{Wcbi}, it follows directly that if the state  
$\hat{\Omega}(t_{\rm d},\theta)$ has no correlations between the battery and the bath, then the contributions to the connection cost are just from the battery passive state. This is, as we previously discussed, because the second term in the last equality in \eqref{Wcbi} vanishes since the battery passive state is diagonal in the eigenbasis of $\hat{H}_{\rm S}$, and therefore ${\rm Tr}_{\rm S}[\hat{Q}_{0}\hat{\rho}_{\rm S}^{\rm p}(t_{\rm d})]=0$. This situation is analogous to the connection cost of the tripartite scenario [see Eq. \eqref{Wctri}], where the first bath is changed to a fresh, thermal second bath. Thus, in the Caldeira-Leggett battery, a difference between the connecting work in the tripartite and bipartite scenario is a direct indication of correlations after disconnection between the battery and the bath.

With the interaction turned on so that the total system Hamiltonian is  \eqref{Hcl}, the battery-bath system autonomously evolves in the charging process with the operator $\hat{U}_{\rm ch}(t)=\exp(-i\hat{H}_{\rm SB}t)$ and, for this scenario to fulfill the cyclic condition it will be necessary that $\hat{\tau}_{\rm MF}=\lim_{t\rightarrow+\infty}{\rm Tr}_{\rm B}[\hat{U}_{\rm ch}(t)\hat{\Omega}(t_{\rm d},\theta) \hat{U}_{\rm ch}^{\dagger}(t)].$ This equality for the Caldeira-Leggett battery is not mathematically guaranteed, but we present strong numerical evidence that, in the thermodynamic limit, it is [Appendix \ref{ApB}].

\subsection{Thermodynamic quantities for the Caldeira-Leggett cycles}

In the tripartite scenario with the Caldeira-Leggett model, a non-trivial consequence of the cyclic property of the battery is the following constraint [see Appendix \ref{ApA}]:
\begin{equation}
{\rm Tr}[\hat{V}_{\rm SB}^{0}\hat{\uptau}]=\lim_{t\rightarrow+\infty}{\rm Tr}[\hat{V}_{\rm SB'}^{0}\hat{\Upomega}(t_{\rm d};t)],
\label{V=V}
\end{equation}
with $\hat{V}_{\rm SB'}^{0}=\hat{V}_{\rm SB}^{0}$. Combining the result in Eq. \eqref{V=V} with the fact that in cycles it holds that ${\rm Tr}[\hat{H}_{\rm S} \hat{\uptau}]=\lim_{t\rightarrow+\infty}{\rm Tr}[\hat{H}_{\rm S} \hat{\Upomega}(t_{\rm d};t)]$, we can conclude from Eq. \eqref{deltaEtri} that
\begin{equation}
    \begin{split}
    W_{\rm diss}(t_{\rm d})&=\lim_{t\rightarrow+\infty}{\rm Tr}[(\hat{H}_{\rm B}+\hat{H}_{\rm B'})\hat{\Upomega}(t_{\rm d};t)]-{\rm Tr}[(\hat{H}_{\rm B}+\hat{H}_{\rm B'})\hat{\uptau}]\\
    &=\Delta E_{\rm B}^{\rm d}(t_{\rm d})+\Delta E_{\rm B'}^{\rm ch}(t_{\rm d}),  
    \end{split}
\label{firstlaw}
\end{equation}
where $\Delta E_{\rm B}^{\rm d}:={\rm Tr}_{\rm B}[\hat{H}_{\rm B}({\rm Tr}_{\rm S}\hat{\rho}_{\rm SB}(t_{\rm d})
-{\rm Tr}_{\rm S}\hat{\tau}_{\rm SB})]$ is the energy change of the first bath in the disconnection process, and $\Delta E_{\rm B'}^{\rm ch}:=\lim_{t\rightarrow+\infty}{\rm Tr}_{\rm B'}[\hat{H}_{\rm B'}({\rm Tr}_{\rm SB}[\hat{\Upomega}(t_{\rm d};t)
]-\hat{\tau}_{\rm B'})]$ is the energy change of the second bath in the charging process, which is negative \cite{hovhannisyan2020}. Note that $\lim_{t\rightarrow+\infty}{\rm Tr}[\hat{H}_{\rm B}\hat{\Upomega}(t_{\rm d};t)]={\rm Tr}_{\rm B}[\hat{H}_{\rm B}\rm Tr_{\rm S}\hat{\rho}_{\rm SB}(t_{\rm d})]$, because the energy of the first bath does not change in the extraction, connection, and charging processes. In particular, in the latter, the bath is not coupled and thus evolves freely. Consequently, from Eq. \eqref{firstlaw}, it follows that any work done by an external agent in the cycle that is not compensated by the extracted ergotropy is directly converted into the sum of the changes in the energies of the first bath in the disconnection process and the second bath in the charging process. This is because Eq. \eqref{V=V} ensures that any energy change of the interaction term between the battery and the baths does not contribute. We recognize Eq. \eqref{firstlaw} as representing the first law of thermodynamics in the cycles. Therefore, following the definition of heat used by the authors in Ref. \cite{esposito2010}, the heat exchanged between the battery and the baths is
\begin{equation}
Q(t_{\rm d}):=-[\Delta E_{\rm B}^{\rm d}(t_{\rm d})+\Delta E_{\rm B'}^{\rm ch}(t_{\rm d})].
\label{heattri}
\end{equation} 
Then, using inequality \eqref{43}, we arrive at the second law for the battery's cycles 
\begin{equation}
    \Sigma(t_{\rm d}):=-\beta Q(t_{\rm d})=D\left(\hat{\Upomega}(t_{\rm d};t)||\hat{\uptau}'\right)\geq0,
    \label{entropyptri}
\end{equation}
where $\Sigma(t_{\rm d})$ is defined as the entropy production of the cycle. This statement is analogous to the classical Clausius statement for a cyclical process in which a system interacts weekly with an infinitely large thermal bath in an isothermal process with inverse temperature $\beta$ \cite{fermi2012,callen1998}.

In analogy to the tripartite case, the cyclic condition for the Caldeira-Leggett battery implies that ${\rm Tr}[\hat{V}_{\rm SB}^{0}\hat{\tau}_{\rm SB}]=\lim_{t\rightarrow+\infty}{\rm Tr}[\hat{V}_{\rm SB}^{0}\hat{\Omega}(t_{\rm d},\theta;t)]$ [see Appendix \ref{ApA}]. Therefore, from Eq. \eqref{WdissBI}, we obtain that the change in energy of the composed system, which is equivalently the dissipative work, is equal to the sum of the change in energy of the bath in the disconnection process and the change in energy of the bath in the charging process, i.e.,
\begin{equation}
\begin{split}
    W_{\rm diss}(t_{\rm d},\theta)&=\Delta E^{\rm cyc}(t_{\rm d},\theta;\infty)\\
    &=\Delta E_{\rm B}^{\rm d}(t_{\rm d})+\Delta E_{\rm B}^{\rm ch}(t_{\rm d},\theta)\\
    &:=\Delta E_{\rm B}^{\rm cyc}(t_{\rm d},\theta),
\end{split}    
    \label{EBbp}
\end{equation}
where $\Delta E_{\rm B}^{\rm d}(t_{\rm d})={\rm Tr}_{\rm B}[\hat{H}_{\rm B}{\rm Tr}_{\rm S}\hat{\rho}_{\rm SB}(t_{\rm d})]-{\rm Tr}_{\rm B}[\hat{H}_{\rm B}{\rm Tr}_{\rm S}\hat{\tau}_{\rm SB}]$ is the energy change of the bath in the disconnection process, and $\Delta E_{\rm B}^{\rm ch}(t_{\rm d},\theta)=\lim_{t\rightarrow+\infty}{\rm Tr}_{\rm B}[\hat{H}_{\rm B}{\rm Tr}_{\rm S}\hat{\Omega}(t_{\rm d},\theta;t)]-{\rm Tr}_{\rm B}[\hat{H}_{\rm B}{\rm Tr}_{\rm S}\hat{\Omega}(t_{\rm d},\theta)]$ is the energy change of the bath in the charging process. In analogy to the tripartite scenario, assuming Eq. \eqref{EBbp} holds, any work done by an external agent in the cycle that is not compensated by the extracted ergotropy is directly converted into a change in the bath energy, $\Delta E_{\rm B}^{\rm cyc}(t_{\rm d},\theta)$. Therefore, similar to Eq. \eqref{firstlaw}, Eq. \eqref{EBbp} represents the first law for each $t_{\rm d}$ and $\theta$ in each cycle. Defining the cycle heat as  $Q(t_{\rm d},\theta):=-\Delta E_{\rm B}^{\rm cyc}(t_{\rm d},\theta)$, we obtain from Eq. \eqref{geq2} a second law analogous to Eq. \eqref{entropyptri}, 
\begin{equation*}
    \Sigma(t_{\rm d},\theta):=-\beta Q(t_{\rm d},\theta)=D(\hat{\Omega}(t_{\rm d},\theta;t)||\hat{\tau}_{\rm SB})\geq 0,
\end{equation*}
where $\Sigma(t_{\rm d},\theta)$ represents the cycles' entropy production.

\section{\label{sec:level1} Numerical implementation}
\subsection{Gaussian quantum mechanics (GQM)}
The systems we have discussed are continuous-variable systems. Furthermore, the applied processes correspond to Gaussian transformations, and the involved states are Gaussian states ~\cite{serafini2017}. For example, the fact that the Hamiltonian of the Caldeira-Leggett model in Eq. \eqref{Hcl} is quadratic in its canonical variables determines that the thermal Gibbs state $\hat{\tau}_{\rm SB}$ in Eq. \eqref{tautri} satisfies the condition of being a particular type of Gaussian state. There is a rich set of mathematical tools~\cite{serafini2017} that allows us to deal effectively with continuous-variable systems in Gaussian states, which we applied in the numerical implementation of previously described cycles with the Caldeira-Leggett model. We present them now, briefly.

When a state of a continuous variable system is Gaussian and its first statistical moments are zero, the relevant quantities that completely characterize it are the second statistical moments. A matrix of second statistical moments, also known as the covariance matrix (CM), of a system with $N$ pairs of canonical operators $(\hat{Q}_{i},\hat{P}_{i})_{i=1}^{N}$ is defined as follow
\begin{equation}
    \mathbf{\sigma}:=\langle\{\mathbf{\hat{r}},\mathbf{\hat{r}}^{\rm T}\}\rangle_{\hat{\rho}},
\end{equation} 
where the symbol $\{\cdot,\cdot\}$ denotes the anti-commutator between pairs of operators $\{\hat{A},\hat{B}\}:=\hat{A}\hat{B}+\hat{B}\hat{A}$, and where the vector of operators $\mathbf{\hat{r}}:=(\hat{Q}_{1},\hat{P}_{1},...,\hat{Q}_{N},\hat{P}_{N})^{\rm T}$ satisfy the canonical commutation relations in the compact form $[\mathbf{\hat{r}},\mathbf{\hat{r}}^{\rm T}]=i\mathbf{\Omega}_{N}$, with $\mathbf{\Omega}_{N}:=\bigoplus
_{i=1}^{N}\begin{pmatrix}
    0&1\\-1&0
\end{pmatrix}$ the \textit{symplectic form} matrix, and the angular brackets represent the trace of a operator of the system with respect to its density operator $\hat{\rho}$ ($\langle\cdot\rangle_{\hat{\rho}}:=\rm Tr[\cdot\hat{\rho}]$). The $\bigoplus$ symbol in $\mathbf{\Omega}_{N}$ denotes the direct sum between matrices. In contrast to the symbol $\rm Tr[\cdot]$, we use the symbol $\rm tr[\cdot]$ to denote traces over matrices.

Now, let us consider only systems with purely quadratic Hamiltonian, i.e., systems with just quadratic combinations of their canonical operators. This Hamiltonian can be written as follow
\begin{equation}
    \hat{H}(t)=\mathbf{\hat{r}}^{\rm T}\mathbf{H}(t)\mathbf{\hat{r}},
    \label{Hquadratic}
\end{equation}
where the variable $t$ represents a possible time dependence within a scalar function and $\mathbf{H}(t)$ is a symmetric and positive semi-definite matrix (also known as the \textit{Hamiltonian matrix} in some mathematical contexts). When a Gaussian state evolves unitarily, its CM evolves as
\begin{equation}
    \mathbf{\sigma}(t)=\mathbf{S}(t)\mathbf{\sigma}(0)\mathbf{S}(t)^{\rm T},
    \label{Sevol}
\end{equation}
where
\begin{equation}
    \mathbf{S}(t)=\hat{\rm T}\exp\left(\int_{0}^{t} \rm d t'2\mathbf{\Omega}_{N}\mathbf{H}(t')\right).
    \label{StT}
\end{equation}
The matrices $\mathbf{S}(t)$ are members of the symplectic group, $Sp(2N,\mathbb{R})$, whose matrices $\mathbf{S}$ satisfy $\mathbf{S}\mathbf{\Omega}_{N}\mathbf{S}^{\rm T}=\mathbf{\Omega}_{N}$ in general. If the Hamiltonian of the system is time-independent, then
\begin{equation}
\mathbf{S}(t)=\exp\left(2\mathbf{\Omega}_{N}\mathbf{H}t\right).
\label{Stind}
\end{equation}

A couple of useful properties of Gaussian states are that the partial trace over the state results in a Gaussian state, i.e., the reduced density operator is Gaussian, and secondly, the mean energy of a Gaussian state can be calculated as 
\begin{equation}
    \frac{1}{2}\rm tr[\mathbf{H}\mathbf{\sigma}],
    \label{EGM}
\end{equation}
as long as the first moments are zero and the Hamiltonian of the system is quadratic like Eq. \eqref{Hquadratic}.
The theoretic aspects of Gaussian states have been thoroughly investigated and applied in mathematical and physical contexts. See Refs. \cite{brask2021,adesso2014,ferraro2005,weedbrook2012,serafini2017,de2006} for comprehensive reviews.

\subsection{Energetic of the cycles in GQM formalism}

The formalism presented above has practical advantages. It allows for the numerical implementation of Gaussian unitary transformations that act on continuous-variable systems with infinite-dimensional Hilbert spaces whose states are Gaussian, which is the case addressed in this work. In this regard, and considering that the state $\hat{\tau}_{\rm SB}$ in Eq. \eqref{tautri} is a Gaussian state with null first moments, for the Caldeira-Leggett model, $W_{\rm d}(t_{\rm d})$ in Eq. \eqref{Wdtri} and $W_{\rm c}(t_{\rm d})$ in Eq. \eqref{Wctri} are expressed as [see Eq. \eqref{EGM}]:
\begin{equation}
    W_{\rm d}(t_{\rm d})=\frac{1}{2}{\rm tr}[\mathbf{H}_{\rm S}\oplus\mathbf{H}_{\rm B}\mathbf{\sigma}_{\rm SB}(t_{\rm d})]-\frac{1}{2}{\rm tr}[\mathbf{H}_{\rm SB}\mathbf{\sigma}_{\rm SB}^{\rm th}]
    \label{WdGM}
\end{equation}
and
\begin{equation}
    W_{\rm c}(t_{\rm d})=\frac{m_0\omega_{\rm R}^{2}}{4}[\mathbf{\sigma}_{\rm S}^{\rm p}(t_{\rm d})]_{11}.
    \label{WcGM}
\end{equation}
The ergotropy is then given by 
\begin{equation}
\begin{split}
    \mathscr{W}(t_{\rm d})&={\rm Tr}_{\rm S}[{\hat{H}_{\rm S}\hat{\rho}_{\rm S}(t_{\rm d})}]-{\rm Tr}_{\rm S}[{\hat{H}_{\rm S}\hat{\rho}_{\rm S}^{\rm p}(t_{\rm d})}]\\
    &=\frac{1}{2}{\rm tr}[\mathbf{H}_{\rm S}\mathbf{\sigma}_{\rm S}(t_{\rm d})]-\frac{1}{2}{\rm tr}[\mathbf{H}_{\rm S}\mathbf{\sigma}_{\rm S}^{\rm p}(t_{\rm d})].
\end{split}
\label{ergoGM}
\end{equation}
Here, $\mathbf{H}_{\rm S}$, $\mathbf{H}_{\rm B}$, and $\mathbf{H}_{\rm SB}$ are the Hamiltonian matrices associated with $\hat{H}_{\rm S}$, $\hat{H}_{\rm B}$, and $\hat{H}_{\rm SB}$ [see Eq. \eqref{Htri}], respectively. $\mathbf{\sigma}_{\rm SB}^{\rm th}$ is the CM that characterizes $\hat{\tau}_{\rm SB}$ [see Eq. \eqref{tautri}], $\mathbf{\sigma}_{\rm SB}(t_{\rm d})=\mathbf{S}_{\rm d}(t_{\rm d})\mathbf{\sigma}_{\rm SB}^{\rm th}\mathbf{S}_{\rm d}(t_{\rm d})^{\rm T}$ is the CM that characterizes $\hat{\rho}_{\rm SB}(t_{\rm d})$, $\mathbf{\sigma}_{\rm S}^{\rm p}(t_{\rm d})$ is the CM that characterizes the passive state $\hat{\rho}_{\rm S}^{\rm p}(t_{\rm d})$, and $\mathbf{\sigma}_{\rm S}(t_{\rm d})$ is the CM that characterizes the reduce state of the battery at time $t_{\rm d}$, $\hat{\rho}_{\rm S}(t_{\rm d})={\rm Tr}_{\rm B}\hat{\rho}_{\rm SB}(t_{\rm d})$. The symplectic evolution $\mathbf{S}_{\rm d}(t_{\rm d})$ can be implemented via numerical approximation of the exact expression of Eq. \eqref{StT}, according to the disconnection protocol $\lambda(t)$ and the Hamiltonian matrix associated with $\hat{H}_{\rm SB}(t)$ in Eq. \eqref{Utrid}. The expressions in Eqs. \eqref{WdGM}, \eqref{WcGM}, and \eqref{ergoGM} must be kept in mind to calculate the energetic of the cycle and the associated quantities from the numerical simulation.

Similarly to the above, for the bipartite scenario, we can write $W_{\rm c}(t_{\rm d},\theta)$ in Eq. \eqref{Wcbi} in this formalism as follows:
\begin{equation}
    W_{\rm c}(t_{\rm d},\theta)=\frac{m_0\omega_{\rm R}^{2}}{4}[\mathbf{\sigma}_{\rm S}^{\rm p}(t_{\rm d})]_{11}-\sum_{k}g_{k}[\mathbf{\sigma}_{\rm SB}(t_{\rm d},\theta)]_{1,2k+1},
\label{WcGMbi}
\end{equation}
where $\mathbf{\sigma}_{\rm SB}(t_{\rm d},\theta)$ is the covariance matrix of the state $\hat{\Omega}(t_{\rm d},\theta)$ in Eq. \eqref{statebitheta} (total system after the extraction). 

In particular, the battery under consideration has only a pair of canonical operators in $\hat{H}_{\rm S}$, $(\hat{Q}_{0},\hat{P}_{0})$, and its possibly active state $\hat{\rho}_{\rm S}(t_{\rm d})$ is a Gaussian state characterized by its covariance matrix $\mathbf{\sigma}_{\rm S}(t_{\rm d})$ (with $2\times2$ dimension). Therefore, its ergotropy \eqref{ergoGM} can be simplified to
\begin{equation}
    \mathscr{W}(t_{\rm d}) = \frac{1}{2}{\rm tr}[\mathbf{\sigma}_{\rm S}(t_{\rm d})\mathbf{H}_{\rm S}]-s(t_{\rm d})\times h,
    \label{ergoG}
\end{equation}
namely, $s(t_{\rm d})$ and $h$ are the only symplectic eigenvalues of $\mathbf{\sigma}_{\rm S}(t_{\rm d})$ and the Hamiltonian matrix $\mathbf{H}_{\rm S}$ (with $2\times2$ dimension), respectively \cite{hovhannisyan2020}. The existence of $s(t_{\rm d})$ and $h$ is guaranteed by the Williamson theorem \cite{simon1999,parthasarathy2012}, and we can find them by calculating the positive eigenvalues of the matrices $i\mathbf{\Omega}_{1}\mathbf{\sigma}_{\rm S}(t_{\rm d})$ and $i\mathbf{\Omega}_{1}\mathbf{H}_{\rm S}$, respectively, where $\mathbf{\Omega}_{1}=\begin{pmatrix}
    0&1\\-1&0
\end{pmatrix}$. See Ref. \cite{hovhannisyan2020} for a detailed derivation of the formula \eqref{ergoG} and notation.

\subsection{Matrices and symplectic operations of the cycles}

Building on our previous discussion, we present the relevant matrices for initializing and executing the cycles in both tripartite and bipartite scenarios. We first construct the thermal covariance matrix $\mathbf{\sigma}_{\rm SB}^{\rm th}$, associated with the thermal state $\hat{\tau}_{\rm SB}$ for the system comprising the battery (central oscillator) and the respective bath. We obtain this by extracting the $N+1$ normal mode frequencies, $\{\tilde{\omega}_j\}_{j=0}^{N}$, from the Hamiltonian matrices $\mathbf{H}_{\rm SB}$ in Eq. \eqref{WdGM} given by the matrix [see Eqs. \eqref{Hcl} and \eqref{Hquadratic}]
\begin{equation}
    \mathbf{H}_{\rm SB}=\frac{1}{2}\begin{pmatrix}m_{0}(\omega_{0}^{2}+\omega_{\rm R}^{2})&0&-g_{1}&0&\cdots&-g_{N}&0\\
    0&1/m_{0}&0&0&\cdots&0&0\\
    -g_{1}&0&m_1\omega_1^{2}&0&\cdots&0&0\\
    0&0&0&1/m_1&\cdots&0&0\\
    \vdots&\vdots&\vdots&\vdots&\ddots&\vdots&\vdots\\
    -g_{N}&0&0&0&\cdots&m_{N}\omega_{N}^{2}&0\\
    0&0&0&0&\cdots&0&1/m_{N}
    \end{pmatrix}.
    \label{Hmatrix}
\end{equation}
Utilizing the $N+1$ normal mode frequencies, we define the Williamson form of $\mathbf{\sigma}^{\rm th}_{\rm SB}$, denoted as $\mathbf{W}_{\mathbf{\sigma}^{\rm th}_{\rm SB}}$, as follows
\begin{equation}
    \mathbf{W}_{\mathbf{\sigma}^{\rm th}_{\rm SB}}=\bigoplus_{j=0}^{N}\begin{pmatrix}\nu_{j}^{\rm th}&0\\
    0&\nu_{j}^{\rm th}
    \end{pmatrix};
    \;\nu_{j}^{\rm th}:=\frac{\exp(\beta\tilde{\omega}_{j})+1}{\exp(\beta\tilde{\omega}_{j})-1}.
    \label{sigmathW}
\end{equation}
Next, we apply the symplectic matrix that transforms the Williamson form into the physics basis. A clear and informative description of this procedure, along with a user-friendly script for constructing $\sigma^{\rm th}_{\rm SB}$, can be found in Ref. \cite{pozas2018}. They also recommend, and we find it highly beneficial, to work with the quadrature base.

In particular for the tripartite case, when the thermal state of the second bath $\rm B'$ is required in the charging process, we construct its covariance matrix using the same structure as the matrix in Eq. \eqref{sigmathW}, but using the natural frequencies in the Hamiltonian $\hat{H}_{\rm B'}=\hat{H}_{\rm B}$ in Eq. \eqref{Hcl}. 

\begin{figure}[h]
\centering
\includegraphics[scale=0.5]{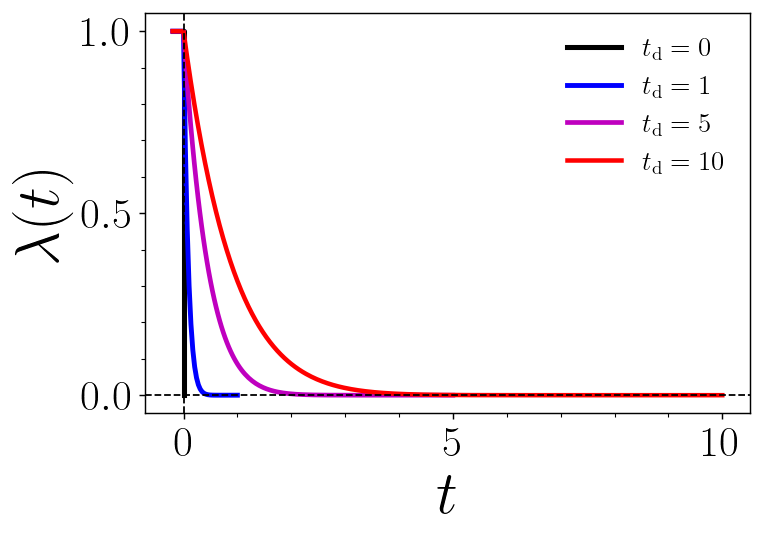}
\caption{Different protocols $\lambda(t)$ as functions of $t$ are shown for various $t_{\rm d}$ values. The black curve, representing a quench-like protocol, corresponds to $t_{\rm d}=0$. For our numerical examples, we employ $\lambda(t)=(1-t/t_{\rm d})^{11}$. This particular exponent is chosen to ensure a similarity to a quench protocol for short disconnection times $t_{\rm d}$, while still allowing the time dependency to influence the battery state.}
\label{fig:protocol}
\end{figure}

We implement the disconnection process via the protocol $\lambda(t)$ [see Fig. \ref{fig:protocol}] for different disconnection times $t_{\rm d}$, following Eqs. \eqref{Sevol} and \eqref{StT}. However, we numerically approximate the symplectic matrix $\mathbf{S}_{\rm d}(t_{\rm d})$, associated to unitary $\hat{U}_{\rm d}(t_{\rm d})$ in Eq. \eqref{Utrid} (which is the same for the bipartite scenario), using the approximation
\begin{equation*}
    \mathbf{S}_{\rm d}(t_{\rm d})\approx\prod_{i=0}^{m}\exp[2\mathbf{\Omega}_{N+1}\mathbf{H}_{\rm SB}[\lambda(i\Delta t)]\Delta t],
\end{equation*}
where the number of iterations $m$ is chosen such that the temporal step $\Delta t:=t_{\rm d}/(m+1)$ is small enough to achieve the desired accuracy. The time-dependent Hamiltonian matrix $\mathbf{H}_{\rm SB}[\lambda(i\Delta t)]$ has the same form as Eq. \eqref{Hmatrix}, but with each $g_{k}$ multiplied by $\lambda(i\Delta t)$ and $\omega_{\rm R}^{2}$ multiplied by $\lambda^{2}(i\Delta t)$. $\mathbf{\Omega}_{N+1}=\bigoplus_{i=1}^{N+1}\begin{pmatrix}
    0&1\\-1&0
\end{pmatrix}$. At the end of the protocol we get the matrix $\mathbf{\sigma}_{\rm SB}(t_{\rm d})=\mathbf{S}_{\rm d}(t_{\rm d})\mathbf{\sigma}^{\rm th}_{\rm SB}(\mathbf{S}_{\rm d}(t_{\rm d}))^{\rm T}$, which is the covariance matrix for the battery and bath at time $t_{\rm d}$. For calculating $W_{\rm d}(t_{\rm d})$ in Eq. \eqref{WdGM}, we must also consider the bare bath Hamiltonian matrix ($\mathbf{H}_{\rm B}$) given by
\begin{equation*}
\mathbf{H}_{\rm B}=\frac{1}{2}
    \begin{pmatrix}
        m_{1}\omega_{1}^{2}&0&\cdots&0\\
        0&1/m_1&\cdots&0\\
        \vdots&\vdots&\ddots&\vdots\\
        0&0&m_{N}\omega_{N}^{2}&0\\
        0&0&0&1/m_{N}
    \end{pmatrix}.
\end{equation*}  
and the bare battery Hamiltonian matrix ($\mathbf{H}_{\rm S}$) given by
\begin{equation}
    \mathbf{H}_{\rm S}=\frac{1}{2}\begin{pmatrix}
        m_{0}\omega_{0}^{2}&0\\0&1/m_{0}
    \end{pmatrix}.
    \label{HSM}
\end{equation}

For the extraction process in the tripartite scenario, we apply by congruence the symplectic matrix $\mathbf{S}_{\mathscr{W}}\oplus\mathbb{I}_{2N\times 2N}$ to the covariance matrix $\mathbf{\sigma}_{\rm SB}(t_{\rm d})$, i.e., we obtain the matrix 
\begin{equation}
\mathbf{\sigma}_{\mathscr{W}}(t_{\rm d}):=(\mathbf{S}_{\mathscr{W}}\oplus\mathbb{I}_{2N\times 2N})\mathbf{\sigma}_{\rm SB}(t_{\rm d})(\mathbf{S}_{\mathscr{W}}\oplus\mathbb{I}_{2N\times 2N})^{\rm T}, 
\label{sigmaergotri}
\end{equation}
which is the covariance matrix of the reduce state ${\rm Tr}_{\rm B'}\hat{\Upomega}_{\rm SB}(t_{\rm d})$ of the tripartite state in Eq. \eqref{tristateendext}.
Here, $\mathbf{S}_{\mathscr{W}}$ is the local symplectic transformation that extracts ergotropy from the battery and is given by
\begin{equation}
    \mathbf{S}_{\mathscr{W}}=-\mathbf{\Omega}_{1}\mathbf{S}_{\mathbf{H}}(\mathbf{S}_{\mathbf{\sigma}})^{\rm T}\mathbf{\Omega}_{1},
    \label{SW}
\end{equation}
where $\mathbf{S}_{\mathbf{H}}$ satisfies $\mathbf{H}_{\rm S}=\mathbf{S}_{\mathbf{H}}\mathbf{W}_{\mathbf{H}_{\rm S}}(\mathbf{S}_{\mathbf{H}})^{\rm T}$, and $\mathbf{S}_{\mathbf{\sigma}}$ satisfies $\mathbf{\sigma}_{\rm S}(t_{\rm d})=\mathbf{S}_{\mathbf{\sigma}}\mathbf{W}_{\mathbf{\sigma}_{\rm S}(t_{\rm d})}(\mathbf{S}_{\mathbf{\sigma}})^{\rm T}$. $\mathbf{W}_{\mathbf{H}_{\rm S}}$ and $\mathbf{W}_{\mathbf{\sigma}_{\rm S}(t_{\rm d})}$ are the Williamson forms of the Hamiltonian matrix $\mathbf{H}_{\rm S}$ in Eq. \eqref{HSM} and $\mathbf{\sigma}_{\rm S}(t_{\rm d})$, respectively. We recommend taking a look at Ref. \cite{hovhannisyan2020}, which provides an explanation of how to obtain the matrix $\mathbf{S}_{\sigma}$ using the Schur decomposition method. $\mathbf{\sigma}_{\rm S}(t_{\rm d})$ corresponds to the first $2\times 2$ block of the $2(N+1)\times2(N+1)$ covariance matrix $\mathbf{\sigma}_{\rm SB}(t_{\rm d})$. See, for example, Eq. \eqref{ergoGM} for the ergotropy in both tripartite and bipartite scenarios, where the covariance matrix $\mathbf{\sigma}_{\rm S}(t_{\rm d})$ explicitly appears. Additionally, in the bipartite scenario, we incorporate the $\theta$-dependent symplectic transformation \cite{adesso2014}
\begin{equation*}
    \mathbf{S}_{\theta}=\begin{pmatrix}
        \cos\theta/2&-\sin\theta/2\\ \sin\theta/2&\cos\theta/2
    \end{pmatrix}
\end{equation*}
associated with the unitary operator $\hat{U}_{\hat{n}_{0}}(\theta)$ in Eq. \eqref{Utheta} along with the matrix $\mathbf{S}_{\mathscr{W}}$. Consequently, the extraction process is realized by the matrix $\mathbf{S}_{\mathscr{W}}(\theta):=(\mathbf{S}_{\theta}\mathbf{S}_{\mathscr{W}})\oplus\mathbb{I}_{2N\times 2N}$, which is applied by congruence to $\mathbf{\sigma}_{\rm SB}(t_{\rm d})$: 
\begin{equation}
\mathbf{\sigma}_{\mathscr{W}}(t_{\rm d},\theta):=\mathbf{S}_{\mathscr{W}}(\theta)\mathbf{\sigma}_{\rm SB}(t_{\rm d})(\mathbf{S}_{\mathscr{W}}(\theta))^{\rm T}.
\label{sigmatdtheta}
\end{equation}
To calculate the connection work $W_{\rm c}(t_{\rm d})$ in Eq. \eqref{WcGM}, it suffices to extract the first diagonal entry of the matrix $\mathbf{\sigma}_{\mathscr{W}}(t_{\rm d})$ \eqref{sigmaergotri}, which corresponds to the $[\mathbf{\sigma}_{\rm S}^{\rm p}(t_{\rm d})]_{11}$ element in Eq. \eqref{WcGM}. In contrast, the calculation of the connection work $W_{\rm c}(t_{\rm d},\theta)$ in Eq. \eqref{WcGMbi} utilizes $\mathbf{\sigma}_{\mathscr{W}}(t_{\rm d},\theta)$. 

Finally, for verifying the expressions in Eqs. \eqref{firstlaw} and \eqref{EBbp} (first laws), the relevant evolution in the charging process, associated with the unitary operator $\hat{U}_{\rm ch}(t)$ in Eq. \eqref{Utrich}, is carried out using the symplectic matrix [see Eq. \eqref{Stind}]
\begin{equation}
    \mathbf{S}_{\rm ch}(t)=\exp(2\mathbf{\Omega}_{N+1}\mathbf{H}_{\rm SB}t).
    \label{SchtGM}
\end{equation}

\subsection{Description of figures}

In the simulations for different disconnection times $t_{\rm d}$, we chose a natural frequency $\omega_{0}=2$ and a mass $m_{0}=1$ for the battery, while the baths --as we said before-- were taken with the same number of oscillators $N=150$ and $m_{k}=1$ (with $k=1,...,N$). The bath frequency sample was generated using the expression in Eq. \eqref{frecuencies} taking $a_{0}=1.03$, and the couplings $g_{k}$'s were generated using the expression in Eq. \eqref{couplings} taking $\gamma=1$ and a cutoff frequency $\omega_{\rm D}=4$. We adjusted $\Delta_{N}$ so as to achieve a good agreement between the discrete value of the normalization frequency, $\omega_{\rm R}^{2}$, in Eq. \eqref{omegaR}, and its value in the continuous limit $\omega_{\rm R}^{2}=2\gamma\omega_{\rm D}$ for the Lorentz-Drude regularization \cite{weiss2012}. The initial temperature in the cycles corresponds to $T=\beta^{-1}=0.1$. As a reference, since we are working in units such that $\hbar=k_{\rm B}=1$, the energy of the thermal state, $\hat{\tau}_{\rm S}=e^{-\beta\hat{H}_{\rm S}}/{\rm Tr}_{\rm S}[e^{-\beta\hat{H}_{\rm S}}]$, of the battery (with $\beta=10$) is ${\rm Tr}_{\rm S}[\hat{H}_{\rm S}\hat{\tau}_{\rm S}]=\frac{\omega_{0}}{2}\coth(\frac{\omega_{0}}{2T})\approx1$.

\begin{figure}[h]
\centering
\includegraphics[scale=0.5]{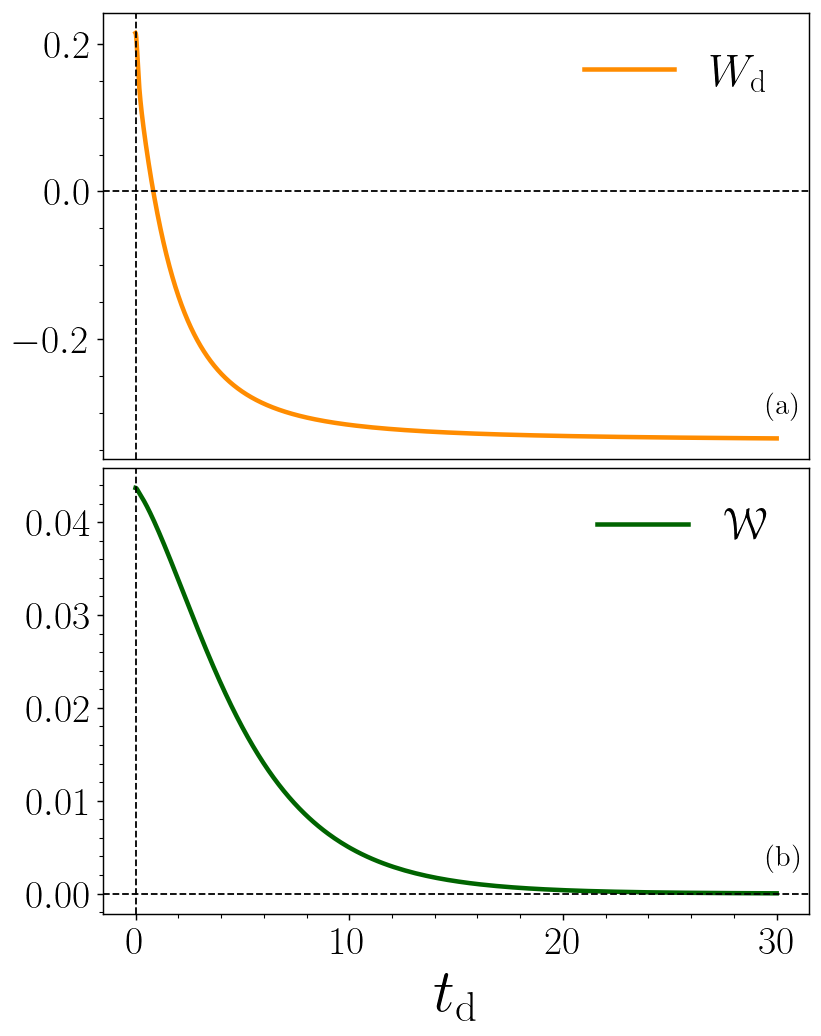}
\caption{Energetic quantities for both tripartite and bipartite scenarios as functions of the disconnection process duration, $t_{\rm d}$. (a) Disconnection work, $W_{\rm d}$. Battery's ergotropy, $\mathrm W$. \textbf{Parameters}: We generate a sample of frequencies defined in Eq. \eqref{frecuencies} with $a_{0}=1.03$ and $N=150$ oscillators in the bath. For the sample of couplings $g_{k}$'s \eqref{couplings}, we take $m_{0}=m_{k}=1$ $\forall k$, $\omega_{\rm D}=4$, and $\gamma=1$. The battery frequency is $\omega_{\rm 0}=2$ and $\beta=10$.}
\label{fig:panel11}
\end{figure}

Fig. \ref{fig:panel11} (a), shows that the disconnection work, $W_{\rm d}(t_{\rm d})$, initially positive for a quench protocol (with $t_{\rm d}=0$), monotonically decreases with increasing disconnection time $t_{\rm d}$ until reaching a negative value where it remains constant. Conversely, Fig. \ref{fig:panel11} (b) shows the ergotropy, $\mathscr{W}(t_{\rm d})$, monotonically decreasing to zero with $t_{\rm d}$, and numerically indistinguishable from it for longer disconnection times. Similarly, the state the battery reaches at prolonged disconnection times is indistinguishable from the thermal state $\hat{\tau}_{\rm S}$, a completely passive state with no ergotropy. In the quench protocol, the ergotropy has its maximum value, $\mathscr{W}_{\rm max}$. In particular, at the end of the process with $t_{\rm d}\sim 4$, $\mathscr{W}$ has decreased by approximately $50\%$ with respect to the quench protocol value $\mathscr{W}_{\rm max}$. By the end of the process with $t_{\rm d}\sim 30$, $\mathscr{W}(t_{\rm d})$ has decreased by approximately $99,9\%$ with respect to $\mathscr{W}_{\rm max}$.

\begin{figure}[h]
\centering
\includegraphics[scale=0.5]{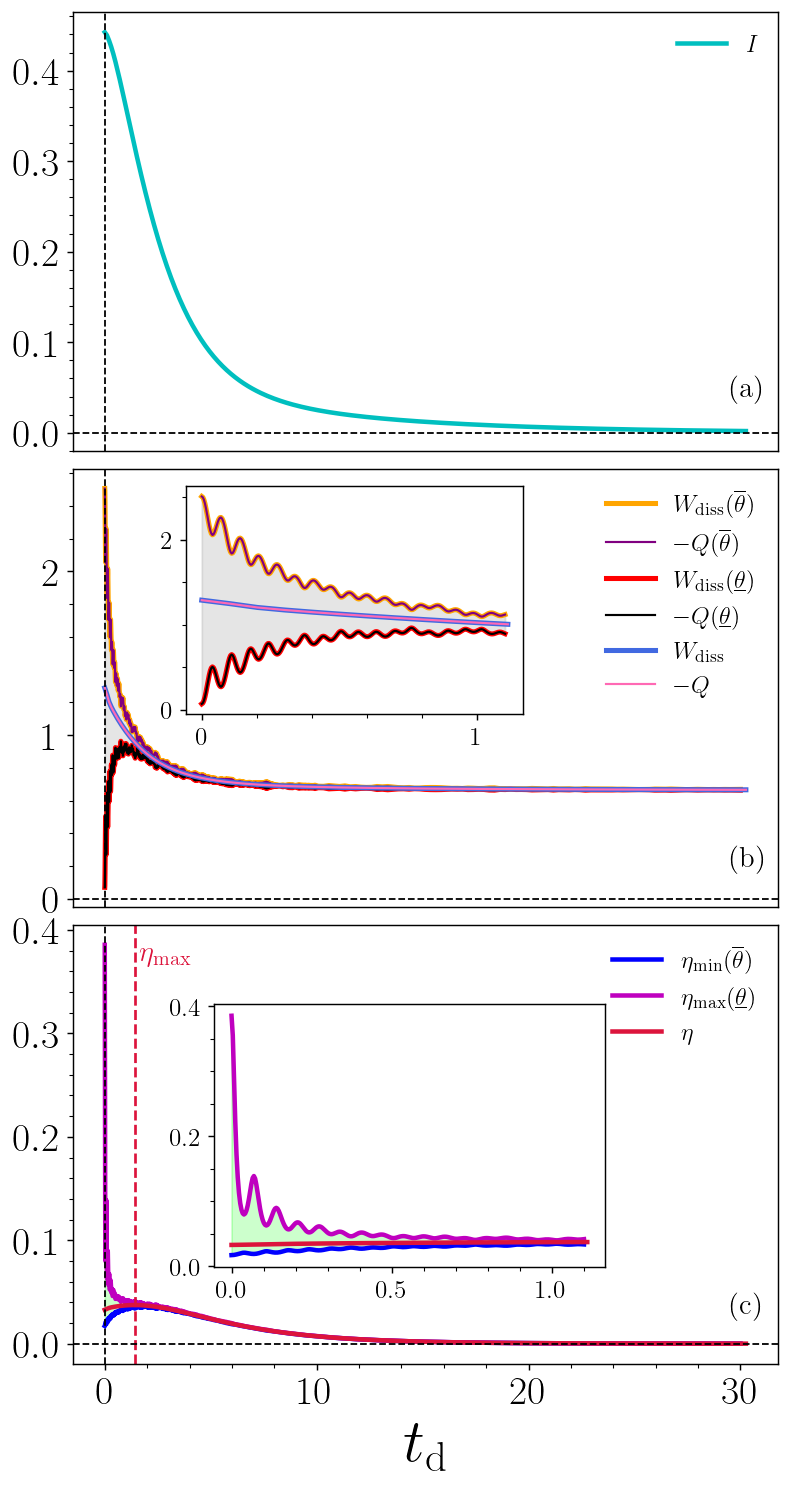}
\caption{The figure depicts energetic quantities and figures of merit for both tripartite and bipartite scenarios as functions of the disconnection process duration, $t_{\rm d}$. (a) Mutual information. (b) First law: Dissipation work and heat. Hatched grey area: Variation of energetic quantities with $\theta$ for a fixed $t_{\rm d}$. (c) Efficiency. Hatched green area: Variation of efficiencies with $\theta$ for a fixed $t_{\rm d}$. Insets in (b) and (c): Zoomed-in view of the main curves at short times. The parameter values are the same as those chosen in Fig. \ref{fig:panel11}, except that in panels (b) and (c), we consider $\theta\in[0,2\pi]$.}
\label{fig:panel21}
\end{figure}

On the other hand, the final correlations between the battery and the respective bath at time $t_{\rm d}$ in the state $\hat{\rho}_{\rm SB}(t_{\rm d})$ [as given by Eq. \eqref{rhoSBtd}] vanish for long disconnection times. This is quantified by the quantum mutual information \cite{stratonovich1965,groisman2005},
\begin{equation} 
I(t_{\rm d})=S[{\rm Tr}_{\rm B}\hat{\rho}_{\rm SB}(t_{\rm d})]+S[{\rm Tr}_{\rm S}\hat{\rho}_{\rm SB}(t_{\rm d})]-S[\hat{\rho}_{\rm SB}(t_{\rm d})],
\label{mutualinformation}
\end{equation} 
which decreases to zero as shown in Fig. \ref{fig:panel21} (a). In other words, if the disconnection protocol takes long enough, the battery and the bath will be completely uncorrelated at the end of the process. In Eq. \eqref{mutualinformation}, $S[\cdot]$ is the von Neumann entropy, which for a Gaussian state with $N$ pairs of canonical operators, can be directly calculated from the $N$ symplectic eigenvalues $\{\nu_{l}\}_{l=1}^{N}$ of the covariance matrix $\mathbf{\sigma}$, of the system state $\hat{\rho}$, by~\cite{serafini2017}
\begin{equation*}
    S(\hat{\rho})=\sum_{l=1}^{N}\left(\frac{\nu_l+1}{2}\right)\ln\left(\frac{\nu_l+1}{2}\right)-\left(\frac{\nu_l-1}{2}\right)\ln\left(\frac{\nu_l-1}{2}\right).
\end{equation*}

Fig. \ref{fig:panel21} (b), shows that in the bipartite case, the dissipated works $W_{\rm diss}(t_{\rm d},\overline{\theta})$ and $W_{\rm diss}(t_{\rm d},\underline{\theta})$ exhibit oscillations as functions of $t_{\rm d}$. Here, $\overline{\theta}$ denotes the values of $\theta$ at which the dissipated work is maximized for a fixed $t_{\rm d}$, while $\underline{\theta}$ represents the values of $\theta$ at which the dissipated work is minimized for a fixed $t_{\rm d}$. The quench protocol yields the maximum value of $W_{\rm diss}(t_{\rm d},\overline{\theta})$, as well as the minimum value of $W_{\rm diss}(t_{\rm d},\underline{\theta})$. As functions of $t_{\rm d}$, $W_{\rm diss}(t_{\rm d},\overline{\theta})$ and $W_{\rm diss}(t_{\rm d},\underline{\theta})$ converge to constant values, with a difference of less than $0.75\%$ between them. This convergence aligns very closely with the convergence of the dissipated work in the tripartite scenario, $W_{\rm diss}(t_{\rm d})$, to long $t_{\rm d}$ times, as evident in the figure. The convergence of the bipartite scenario to the value of the dissipated work of the tripartite scenario is justified by the fact that for disconnection protocols of prolonged duration, at the end of the process, the correlations between the battery and the bath are lost, as can be clearly seen in the behavior of the mutual information, $I(t_{\rm d})$, in panel (a). Therefore, the connection work (on which the dissipated work depends) would be determined only by the passive state of the battery, as can be deduced from Eq. \eqref{Wcbi}.

It's important to note that the dissipated work always remains positive regardless of the scenario, confirming the findings of Eqs. \eqref{43} and \eqref{geq2}. Additionally, excellent agreement exists between $W_{\rm diss}(t_{\rm d},\overline{\theta})$ and $-Q(t_{\rm d},\overline{\theta})$, and similarly between  $W_{\rm diss}(t_{\rm d},\underline{\theta})$ and $-Q(t_{\rm d},\underline{\theta})$ as stated by the first laws in Eq. \eqref{EBbp}. We remind the reader that this is derived from the equality in Appendix \ref{ApA} under the cycle condition, which is shown to be valid in a broad parameter regime in Appendix \ref{ApB}. It is worth highlighting this last point since the $\theta$-dependence of the dissipated work is contained in the connection work and comes from the correlations between the battery and the bath after the extraction of the ergotropy. Conversely, the $\theta$-dependence of the heat is contained in the state of the bath at the end of the charging process, even though the initial state of the bath was independent of $\theta$, as we remarked previously in Eq. \eqref{ec42}. Finally, the agreement between $W_{\rm diss}(t_{\rm d})$ and $-Q(t_{\rm d})$ in the tripartite cases is as expected according to Eqs. \eqref{firstlaw} and \eqref{heattri}, provided that the bath is considered to be close (numerically) to the thermodynamic limit. 

The most interesting results of this work are found in Fig. \ref{fig:panel21} (c). Here, we observe that the tripartite scenario's efficiency, $\eta(t_{\rm d})$ in  Eq. \eqref{etatri}, presents a maximum, $\eta_{\rm max}$, at $t_{\rm d}\sim 1.4$, which is approximately $12\%$ greater than the value for a quench protocol, $\eta(t_{\rm d}=0)$. Thus, although the ergotropy is maximized for the quench protocol, efficiency is not. Therefore, delaying the disconnection for greater energy benefit is advantageous.

Fig. \ref{fig:panel21} (c) also shows that in the bipartite scenario, the efficiency, $\eta_{\rm max}(t_{\rm d},\underline{\theta})$; which corresponds to the maximum values of the efficiency, $\eta(t_{\rm d},\theta)$ in Eq. \eqref{etabi}, in which the dissipated work $W_{\rm diss}(t_{\rm d},\theta)$ is minimum as a function of $\theta$ for fixed $t_{\rm d}$, presents local maxima that decay in amplitude as $t_{\rm d}$ increases [see panel (b)]. Its maximum value corresponds to the quench protocol, $\eta_{\rm max}(t_{\rm d}=0,\underline{\theta})$, which is one order of magnitude greater than the maximum of the efficiency of the tripartite case ($\eta_{\rm max}$). Similarly, $\eta_{\rm min}(t_{\rm d},\overline{\theta})$ (for values of $\overline{\theta}$ at which $W_{\rm diss}(t_{\rm d},\theta)$ is maximum), also presents fluctuations. Its minimum value is achieved in the quench protocol, $\eta_{\rm min}(t_{\rm d}=0,\overline{\theta})$, which differs by about $ 46\%$ from the value of $\eta_{\rm max}$. This demonstrates that utilizing the bipartite scenario and the battery-bath correlations in ultra-fast disconnection processes for charging and discharging is considerably more advantageous than the tripartite case, where refreshing the bath eliminates these correlations. However, without an accurate control of the value of $\theta$, the efficiency becomes highly fluctuating for these ultra-fast processes. 

In summary, the tripartite case offers the benefit of exploiting the limited time constraint by finding the optimal protocol and $t_{\rm d}$ for higher efficiency if a quench protocol is not technologically feasible or the variable $\theta$ is not under control. Conversely, the bipartite scenario allows exploiting the battery-bath correlations to achieve efficiencies surpassing the tripartite case significantly by minimizing disconnection time. 

Finally, it is worth noting that both scenarios exhibit zero efficiencies for long disconnection protocols, rendering the battery useless as the ergotropy also becomes zero. This battery is a useful device only if operated out of equilibrium and in the strong coupling regime.

\section{summary and conclusions}\label{sec:conclu}

Let us summarize our main results. We studied a simple model for a quantum battery charged by thermalization using the Caldeira-Leggett model for a harmonic oscillator in the role of the battery. In essence, one considers a cycle in which the battery is disconnected from the bath, its ergotropy is extracted, and then the battery is recharged by bringing it back in contact with the thermal bath. This setup was previously studied using a sudden disconnecting process.
In this study, we considered the situation in which the disconnecting process lasts a time $t_{\rm d}$ and found that the ergotropy of the disconnected battery and the work required to disconnect it decreases as $t_{\rm d}$ increases. These results are expected because if the disconnecting time $t_{\rm d}$ is long, the process is quasi-static, and at every moment, the battery is in equilibrium with the bath. At a long $t_{\rm d}$, its reduced state will be the completely passive Gibbs state (with vanishing ergotropy). Thus, the efficiency is expected to be zero if $t_{\rm d}$ is long, and the battery is efficient only for short dissipative processes. Thus, the main object of our analysis was the efficiency (the ratio of ergotropy to the sum of the disconnecting and connecting work) as a function of $t_{\rm d}$. We considered two scenarios. In the first, the battery is reconnected to a fresh thermal bath at the same temperature as the initial system. We observed an optimal disconnection time at which the efficiency of the battery peaks, meaning that initially, ergotropy decreases slower than the disconnecting work. However, it is possible to consider a second scenario. In the second scenario, the battery is reconnected to the same bath from which it was disconnected. Because the battery and the bath are correlated, the connecting work depends on the process that extracted the ergotropy of the battery. This effect is not present in the first scenario. In this case, we look for optimal disconnecting times and optimal ergotropy extraction processes. We found that the highest efficiency is achieved with $t_{\rm d}=0$. Still, at these short times, the efficiency sensitivity to the ergotropy extraction process (i.e., to $\theta$) is very high. As $t_{\rm d}$ increases, this sensitivity decreases, and the efficiency becomes almost the same as in the first scenario. This can be linked to the observed decrease of correlations as $t_{\rm d}$ increases.

Apart from the results above, we have also explored the thermodynamics of these cycles. While for the first scenario, there is rigorous proof that the battery converges to the initial state, the mean-force Gibbs state, this is not so for the second scenario. We have shown numerically that in this second scenario, this condition holds [Appendix \ref{ApB}]. The cyclic condition has an important consequence for the Caldeira-Leggett model, which is an equality proved in Appendix \ref{ApA}. The thermodynamic consequence of that equality is this: the total work performed in the cycle equals the energy change of the bath in the cycle, i.e., $W(\theta)+Q(\theta)=0$, allowing us to express the second law for the cycle as in the standard Clausius statement $Q(\theta)\leq 0$. This looks hardly surprising because it is the expected result for systems weakly coupled to a bath. But for systems strongly coupled to the bath, there is debate on what quantities constitute the heath, entropy production, etc. For example, see Refs~\cite{esposito2010,manzano2016,bera2017,strasberg2019,strasberg2020,dolatkhah2020,landi2021,riechers2021,elouard2023} which adopt different formalisms for thermodynamics in situations involving correlations, strong coupling, or interaction with non-equilibrium or thermal baths. Here, heat is the energy change of the bath like in Esposito et al.'s definition~\cite{esposito2010}. 

It is interesting to explore if these results can be extended to other system-bath couplings. Positive results point to a possible unification of the thermodynamics of cycles in the strong and weak coupling.

Finally, let us mention that although we have shown numerical results for a given disconnecting protocol with various $t_{\rm d}$, we have observed the same qualitative results with various other time-dependent functions $\lambda(t)$ that monotonically decrease from $\lambda(0)=1$ to $\lambda(t_{\rm d})=0$.

\begin{acknowledgments}
F. B. thanks Fondecyt project 1231210 and the Millennium Nucleus “Physics of active
matter” of ANID (Chile). D. F. acknowledges partial support from Fondecyt project 1231210.
\end{acknowledgments}

\appendix

\section{\label{ApB} Bipartite scenario. Evolution of the battery covariance matrix during the charging process and convergence to the mean-force Gibbs state.}

\begin{figure}[h]
\centering
\begin{subfigure}{0.45\textwidth}
\includegraphics[width=\linewidth]{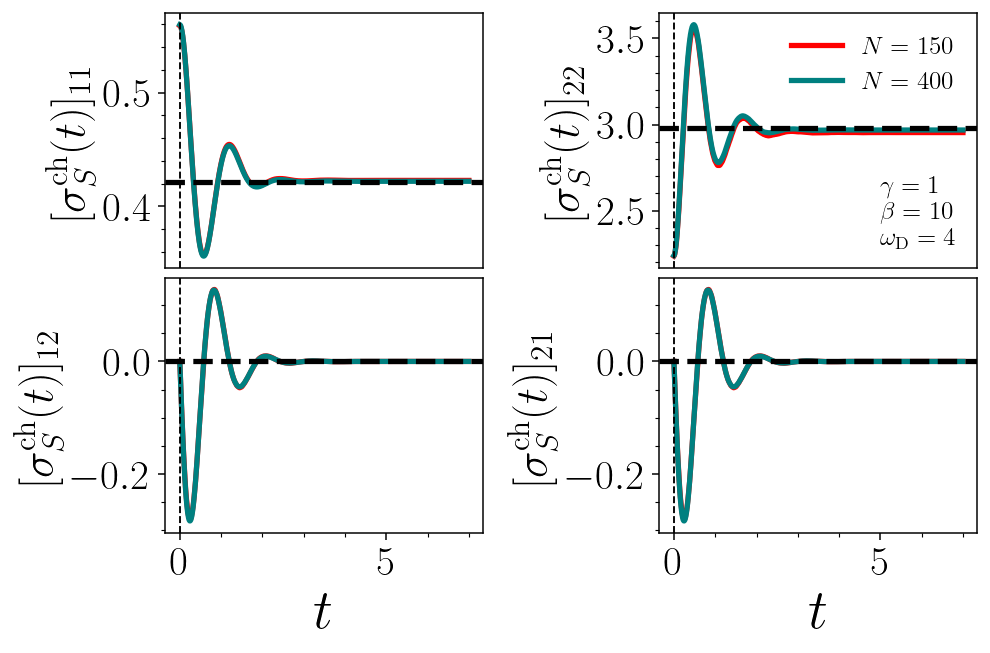}
\end{subfigure}
\begin{subfigure}{0.45\textwidth}
\includegraphics[width=\linewidth]{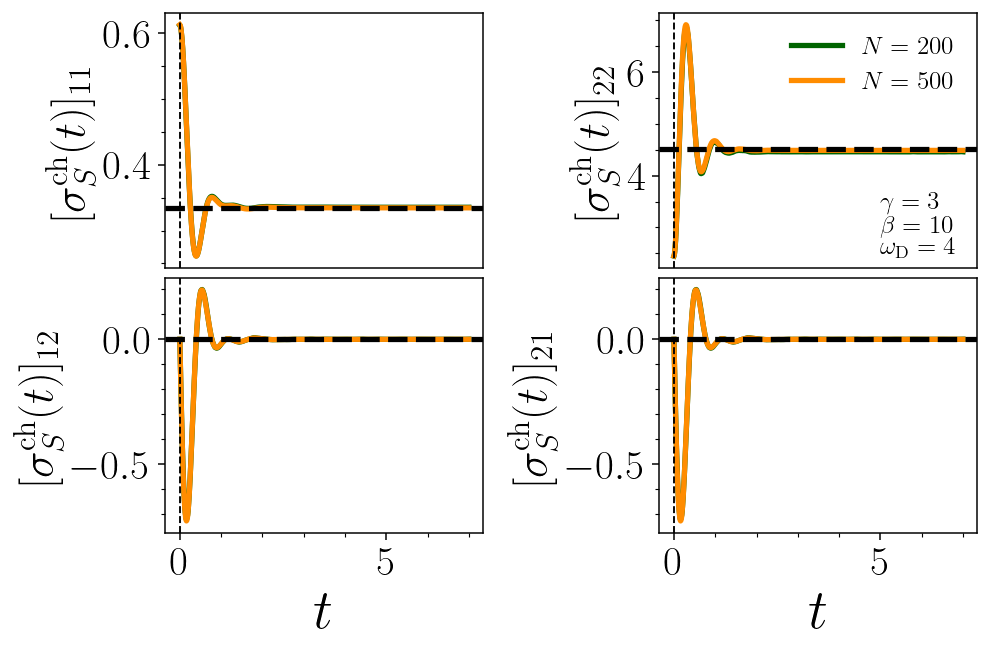}
\end{subfigure}
\begin{subfigure}{0.45\textwidth}
\includegraphics[width=\linewidth]{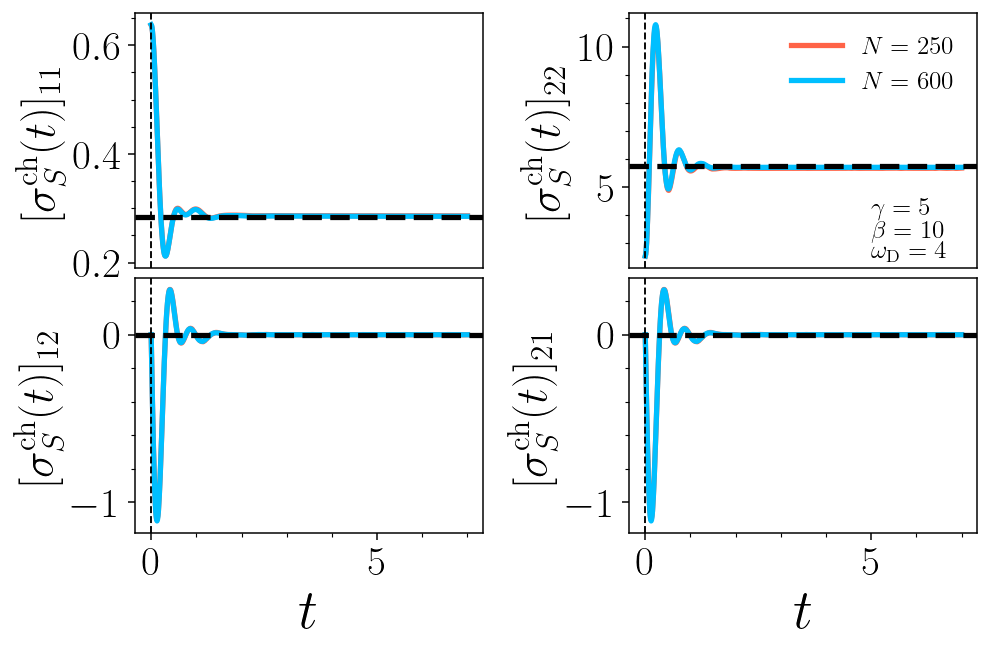}
\end{subfigure}
\begin{subfigure}{0.45\textwidth}
\includegraphics[width=\linewidth]{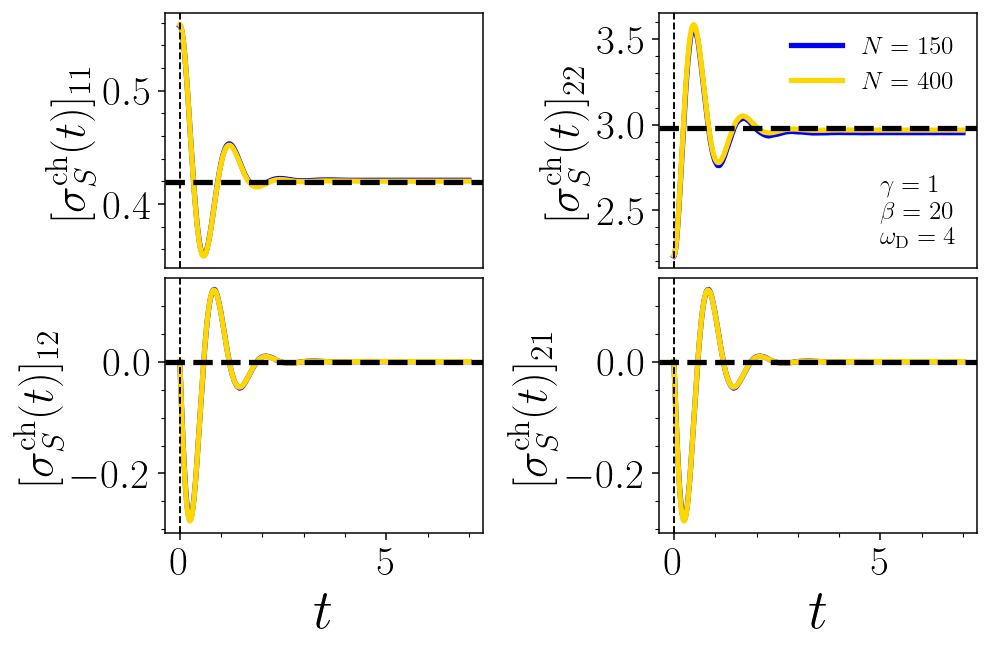}
\end{subfigure}
\begin{subfigure}{0.45\textwidth}
\includegraphics[width=\linewidth]{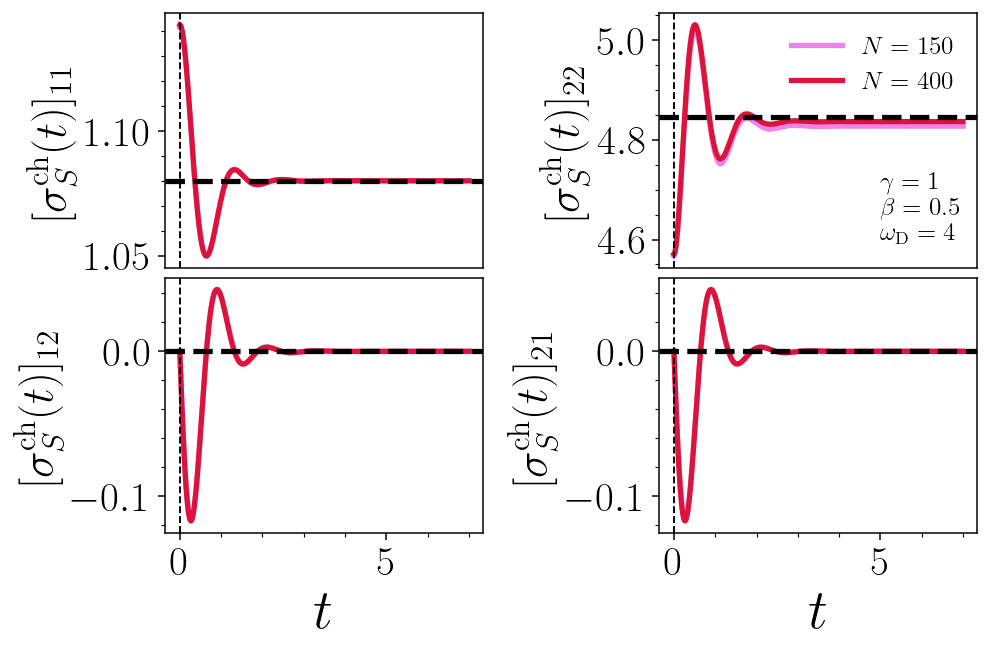}
\end{subfigure}
\begin{subfigure}{0.45\textwidth}
\includegraphics[width=\linewidth]{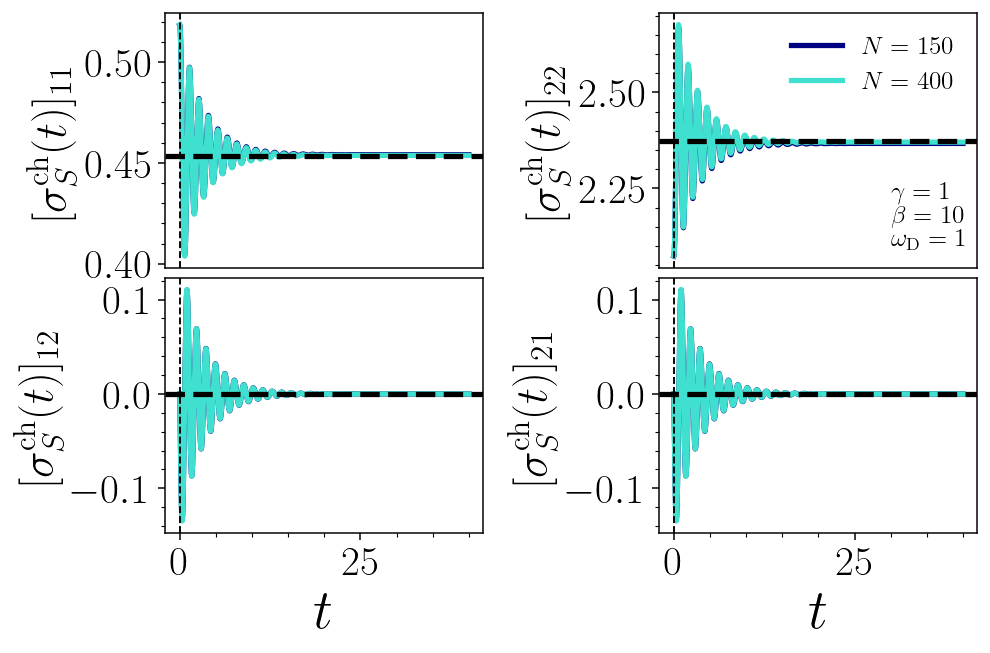}
\end{subfigure}
\caption{Bipartite scenario. Evolution of the battery covariance matrix, $\mathbf{\sigma}_{S}^{\rm ch}(t)$, in the charging process for different values of $N$, $\gamma$, $\beta$ and $\omega_{\rm D}$. In each case, the initial CM of the charging process corresponds to $\mathbf{\sigma}_{\mathscr{W}}(t_{\rm d}=0)$ in Eq. \eqref{Omegaquench}. The values of the relevant parameters influencing convergence (horizontal dashed lines) to the mean-force Gibbs state [Eqs. \eqref{Q2stationary} and \eqref{P2stationary}] are specified in each panel, whereas $\omega_{0}=2$ and $m_{0}=1$. The bath frequency sampling and the coupling sample are generated using expressions in Eqs. \eqref{couplings} and \eqref{frecuencies}.}
\label{fig:panels}
\end{figure}

In Fig. \ref{fig:panels}, we show the evolution of the battery covariance matrix, $\mathbf{\sigma}_{S}^{\rm ch}(t)$, during the charging process for the bipartite case. We take the initial state of the process to be the state with covariance matrix
\begin{equation}
\mathbf{\sigma}_{\mathscr{W}}(t_{\rm d}=0):=(\mathbf{S}_{\mathscr{W}}\oplus\mathbb{I}_{2N\times2N})\mathbf{\sigma}_{\rm SB}^{\rm th}(\mathbf{S}_{\mathscr{W}}\oplus\mathbb{I}_{2N\times2N})^{\rm T},
\label{Omegaquench}
\end{equation}
which is the result of preparing the battery and bath in the state $\hat{\tau}_{\rm SB}$ (with CM $\mathbf{\sigma}_{\rm SB}^{\rm th}$), then disconnecting them via a quench, and finally extracting the ergotropy with the operator $\hat{U}_{\mathscr{W}}$ in Eq. \eqref{UW}, equivalently, with the symplectic transformation $\mathbf{S}_{\mathscr{W}}$ in Eq. \eqref{SW}. In the charging process, $\mathbf{\sigma}_{\mathscr{W}}(t_{\rm d}=0)$ evolves by the symplectic matrix $\mathbf{S}_{\rm ch}(t)$ in Eq. \eqref{SchtGM}. We choose different values for the parameters $N$, $\gamma$, $\beta$, and $\omega_{\rm D}$ for each panel, while the parameters associated with the battery are kept the same. Evolving the battery from the composite state $\mathbf{\sigma}_{\mathscr{W}}(t_{\rm d}=0)$ is the more relevant to show that the battery satisfies a cycle in the bipartite scenario, because it is the state that presents the highest battery-bath correlations, as can be seen in Fig. \ref{fig:panel21} (a), and because numerically we corroborate that it is the state in which the reduced bath state is the least thermal [unlike any state $\mathbf{\sigma}_{\mathscr{W}}(t_{\rm d}\neq0,\theta=0)$, see Eq. \eqref{sigmatdtheta}]. From the figure, we can observe that in all the cases considered, the battery tends to converge better to the mean-force Gibbs state as $N$ increases.

\newpage
\section{\label{ApA} Calculation to obtain the identity in Eq. \eqref{V=V}.}

Here we present the calculations behind the identity
\begin{equation*}
{\rm Tr}[\hat{V}_{\rm SB}^{0}\hat{\uptau}]=\lim_{t\rightarrow+\infty}{\rm Tr}[\hat{V}_{\rm SB'}^{0}\hat{\Upomega}(t_{\rm d};t)],
\end{equation*}
which holds in the thermodynamic limit. We recall that we are considering the same size, the same masses, the same frequency spectrum, and the same couplings for both baths $\rm B$ and $\rm B'$. On one hand, for the right-hand side, using Eqs. \eqref{Uchbi}, \eqref{Omegatri} and \eqref{Utrich}, we have that
\begin{equation}
\begin{split}
\lim_{t\rightarrow+\infty}{\rm Tr}[\hat{V}_{\rm SB'}^{0}\hat{\Upomega}(t_{\rm d};t)]&\overset{(\S 1)}{=}\lim_{t\rightarrow+\infty}{\rm Tr}\left[\hat{\mathcal{U}}_{\rm ch}^{\dagger}(t)\hat{V}_{\rm SB'}^{0}\hat{\mathcal{U}}_{\rm ch}(t)\hat{\Upomega}(t_{\rm d})\right]\\
&\overset{(\S 2)}{=}\lim_{t\rightarrow+\infty}{\rm Tr}\left[\hat{U}_{\rm ch}^{\dagger}(t)\hat{V}_{\rm SB'}^{0}\hat{U}_{\rm ch}(t)\hat{\Upomega}(t_{\rm d})\right]\\
&\overset{(\S 3)}{=}\lim_{t\rightarrow+\infty}{\rm Tr}_{\rm S,B'}\left[\left(\frac{m_{0}\omega_{\rm R}^{2}}{2}\hat{Q}_{0}^{2}(t)-\hat{Q}_{0}(t)\sum_{k}g_{k}\hat{Q}_{k}(t)\right)\hat{\rho}_{\rm S}^{\rm p}(t_{\rm d})\otimes\hat{\tau}_{\rm B'}\right]\\
&\overset{(\S 4)}{=}\lim_{t\rightarrow+\infty}{\rm Tr}_{\rm S,B'}\left[\left(\frac{m_{0}\omega_{\rm R}^{2}}{2}\hat{Q}_{0}^{2}(t)-m_{0}\hat{Q}_{0}(t)\left[\frac{\ud^{2}\hat{Q}_{0}(t)}{\ud t^{2}}+(\omega_{0}^{2}+\omega_{\rm R}^{2})\hat{Q}_{0}(t)\right]\right)\hat{\rho}_{\rm S}^{\rm p}(t_{\rm d})\otimes\hat{\tau}_{\rm B'}\right]\\
&\overset{(\S 5)}{=}-m_{0}\left(\omega_{0}^{2}+\frac{\omega_{\rm R}^{2}}{2}\right)\lim_{t\rightarrow+\infty}{\rm Tr}_{\rm S,B'}\left[\hat{Q}_{0}^{2}\hat{U}_{\rm ch}(t)\hat{\rho}_{\rm S}^{\rm p}(t_{\rm d})\otimes\hat{\tau}_{\rm B'}\hat{U}_{\rm ch}(t)^{\dagger}\right]\\
&-m_{0}\lim_{t\rightarrow+\infty}{\rm Tr}_{\rm S,B'}\left[\hat{Q}_{0}(t)\frac{\ud^{2}\hat{Q}_{0}(t)}{\ud t^{2}}\hat{\rho}_{\rm S}^{\rm p}(t_{\rm d})\otimes\hat{\tau}_{\rm B'}\right]\\
&\overset{(\S6)}{=}-m_{0}\left(\omega_{0}^{2}+\frac{\omega^{2}_{\rm R}}{2}\right){\rm Tr}_{\rm S}[\hat{Q}_{0}^{2}\hat{\tau}_{\rm MF}]+\frac{1}{m_{0}}\lim_{t\rightarrow+\infty}{\rm Tr}_{\rm S,B'}[\hat{P}_{0}^{2}\hat{U}_{\rm ch}(t)\hat{\rho}_{\rm S}^{\rm p}(t_{\rm d})\otimes\hat{\tau}_{\rm B'}\hat{U}_{\rm ch}(t)^{\dagger}]\\
&\overset{(\S 7)}{=}-m_{0}\left(\omega_{0}^{2}+\frac{\omega^{2}_{\rm R}}{2}\right){\rm Tr}_{\rm S}[\hat{Q}_{0}^{2}\hat{\tau}_{\rm MF}]+\frac{1}{m_0}{\rm Tr}_{\rm S}[\hat{P}_{0}^{2}\hat{\tau}_{\rm MF}],
\end{split}
\label{Vfinaltri}
\end{equation}
where in the equality ($\S 1$), we use the cyclic property of the trace. In equality ($\S 2$), we use the fact that $[\exp(-i\hat{H}_{\rm B}t),\hat{V}^{0}_{\rm SB'}]=0$. In equality ($\S 3$), we trace out the first bath's degrees of freedom and express the operators in the Heisenberg picture, where the operators $\hat{Q}_{0}(t)$ and $\hat{P}_{0}(t)$ satisfy the Heisenberg equations
\begin{equation}
\begin{split}
    \frac{\ud \hat{Q}_{0}(t)}{\ud t}&=-\frac{i}{\hbar}[\hat{Q}_{0}(t),\hat{H}_{\rm SB'}]\\
    &=\frac{\hat{P}_{0}(t)}{m_{0}}
\end{split}
\label{Qheisenberg}
\end{equation}
and
\begin{equation}
\begin{split}
    \frac{\ud \hat{P}_{0}(t)}{\ud t}&=-\frac{i}{\hbar}[\hat{P}_{0}(t),\hat{H}_{\rm SB'}]\\
    &=-m_{0}(\omega_{0}^{2}+\omega_{\rm R}^{2})\hat{Q}_{0}(t)+\sum_{k}g_{k}\hat{Q}_{k}(t),
    \label{Pheisenberg}
\end{split}
\end{equation}
respectively. In equality ($\S 4$), we replace the following equality:
\begin{equation}
    m_0\left[\frac{\ud ^{2}\hat{Q}_{0}(t)}{\ud t^{2}}+(\omega_{0}^{2}+\omega_{\rm R}^{2})\hat{Q}_{0}\right]=\sum_{k}g_{k}\hat{Q}_{k},
    \label{Eqd2Q}
\end{equation}
which is obtained using the Heisenberg equations \eqref{Qheisenberg} and \eqref{Pheisenberg}. In equality ($\S 6$), we use the equality
\begin{equation}
\lim_{t\rightarrow+\infty}{\rm Tr}_{\rm S,B'}[\hat{Q}_{0}(t)\frac{\ud^{2}\hat{Q}_{0}(t)}{\ud t^{2}}\hat{\rho}_{\rm S}^{\rm p}(t_{\rm d})\otimes\hat{\tau}_{\rm B'}]\\=-\lim_{t\rightarrow+\infty}{\rm Tr}_{\rm S,B'}[\frac{\hat{P}_{0}^{2}(t)}{m_{0}^{2}}\hat{\rho}_{\rm S}^{\rm p}(t_{\rm d})\otimes\hat{\tau}_{\rm B'}],
\label{barraident}
\end{equation}
which is valid in the thermodynamic limit, where the battery satisfies the Langevin equation 
\begin{equation*}
    \dv[2]{\hat{Q}_{0}(t)}{t}+\int_{0}^{t}\ud t'\gamma(t-t')\dv{\hat{Q}_{0}(t')}{t'}\\+\omega_{0}^{2}\hat{Q}_{0}(t)=\frac{\hat{\xi}(t)}{m_0}-\gamma^{(2)}(t)\hat{Q}_{0}(0),
\end{equation*}
which is obtained using equations \eqref{Qheisenberg} and \eqref{Pheisenberg} and solving the Heisenberg equation \cite{weiss2012}
\begin{equation*}
    \begin{split}
        \frac{\ud^{2}\hat{Q}_{k}(t)}{\ud t^{2}}=\omega_{k}^{2}\hat{Q}_{k}(t)+\frac{g_{k}}{m_{k}}\hat{Q}_{0}(t).
    \end{split}
\end{equation*}
$\gamma(t)$ and $\hat{\xi}(t)$ are defined as the memory kernel \eqref{gamma} and the noise operator \eqref{noise}, respectively, but with the operator and parameters of the second bath $\rm B'$ and respective couplings. See Ref. \cite{hovhannisyan2020} for a detailed derivation of Eq. \eqref{barraident}. In equality ($\S 7$), we trace out the second bath's degrees of freedom, obtaining the mean-force Gibbs state at long-time behavior. 

On the other hand, for the left-hand side, using Eq. \eqref{tautri}, we have that 
\begin{equation}
\begin{split}
{\rm Tr}[\hat{V}_{\rm SB}^{0}\hat{\uptau}]&\overset{(\S\S1)}{=}{\rm Tr}_{\rm S,B}\left[\hat{V}_{\rm SB}^{0}\hat{\tau}_{\rm SB}\right]\\
&\overset{(\S\S2)}{=}{\rm Tr}_{\rm S,B}\left[\hat{V}_{\rm SB}^{0}\exp(-i\hat{H}_{\rm SB}t)\exp(i\hat{H}_{\rm SB}t)\hat{\tau}_{\rm SB}\right]\\
&\overset{(\S\S3)}{=}{\rm Tr}_{\rm S,B}\left[\exp(i\hat{H}_{\rm SB}t)\hat{V}_{\rm SB}^{0}\exp(-i\hat{H}_{\rm SB}t)\hat{\tau}_{\rm SB}\right]\\
&\overset{(\S\S4)}{=}{\rm Tr}_{\rm S,B}\left[\left(\frac{m_{0}\omega_{\rm R}^{2}}{2}\hat{Q}_{0}(t)-\hat{Q}_{0}(t)\sum_{k}g_{k}\hat{Q}_{k}(t)\right)\hat{\tau}_{\rm SB}\right]\\
&\overset{(\S\S5)}{=}{\rm Tr}_{\rm S,B}\left[\left(\frac{m_{0}\omega_{\rm R}^{2}}{2}\hat{Q}_{0}(t)-m_{0}\hat{Q}_{0}(t)[\frac{\ud^2\hat{Q}_{0}(t)}{\ud t^{2}}+(\omega_{0}^{2}+\omega_{\rm R}^{2})\hat{Q}_{0}(t)]\right)\hat{\tau}_{\rm SB}\right]\\
&\overset{(\S\S6)}{=}-{\rm Tr}_{\rm S,B}\left[\left(m_0[\omega_{0}^{2}+\frac{\omega_{\rm R}^{2}}{2}]\hat{Q}_{0}^{2}(t)-\frac{\hat{P}_{0}^{2}(t)}{m_{0}}+\frac{\ud [\hat{Q}_{0}(t)\hat{P}_{0}(t)]}{\ud t}\right)\hat{\tau}_{\rm SB}\right]\\
&\overset{(\S\S7)}{=}-m_{0}\left(\omega_{0}^{2}+\frac{\omega^{2}_{\rm R}}{2}\right){\rm Tr}_{\rm S,B}[\hat{Q}_{0}^{2}\hat{\tau}_{\rm SB}]+\frac{1}{m_{0}}{\rm Tr}_{\rm S,B}[\hat{P}_{0}^{2}\hat{\tau}_{\rm SB}]\\
&\overset{(\S\S8)}{=}-m_{0}\left(\omega_{0}^{2}+\frac{\omega^{2}_{\rm R}}{2}\right){\rm Tr}_{\rm S}[\hat{Q}_{0}^{2}\hat{\tau}_{\rm MF}]+\frac{1}{m_0}{\rm Tr}_{\rm S}[\hat{P}_{0}^{2}\hat{\tau}_{\rm MF}],
\end{split}
\label{Vinitialtri}
\end{equation}
where in equality ($\S\S 1$),  we trace out the second bath's degrees of freedom. In equality ($\S\S 2$), we incorporate the identity operator $\hat{\mathbb{I}}_{\rm SB}=\exp(-i\hat{H}_{\rm SB}t)\exp(i\hat{H}_{\rm SB}t)$ on the Hilbert space of the subsystem $\rm S\cup B$. In equality ($\S\S 3$), we use the fact that $[\hat{\tau}_{\rm SB},\exp(-i\hat{H}_{\rm SB}t)]=0$ and the cyclic property of the trace. In equality ($\S\S 4$), we express the operators in the Heisenberg picture, where the operators $\hat{Q}_{0}(t)$ and $\hat{P}_{0}(t)$ satisfy the Heisenberg equations \eqref{Qheisenberg} and \eqref{Qheisenberg}, respectively. In equality ($\S\S 5$), we replace the equality \eqref{Eqd2Q}. In equality ($\S\S 6$), we use the product rule and Eq. \eqref{Qheisenberg} to obtain that
\begin{equation*}
    \hat{Q}_{0}(t)\frac{\ud^{2}Q_{0}(t)}{\ud t^{2}}=-\frac{\hat{P}_{0}^{2}}{m_{0}^{2}}+\frac{1}{m_{0}}\frac{\ud (\hat{Q}_{0}(t)\hat{P}_{0}(t))}{\ud t},
\end{equation*}
which we then replace. In equality ($\S\S 7$), we use the fact that  $\ud {\Tr_{\rm S,B}}[\hat{Q}_{0}(t)\hat{P}_{0}(t)\hat{\tau}_{\rm SB}]/\ud t=0$, and, finally, we trace out the first bath's degrees of freedom in equality ($\S\S8$).

In analogy to the tripartite case, the cyclic condition for the Caldeira-Leggett battery in the bipartite scenario implies that ${\rm Tr}[\hat{V}_{\rm SB}^{0}\hat{\tau}_{\rm SB}]=\lim_{t\rightarrow+\infty}{\rm Tr}[\hat{V}_{\rm SB}^{0}\hat{\Omega}(t_{\rm d},\theta;t)]$. This equivalence can be established using the same methodology employed for Eqs. \eqref{Vfinaltri} and \eqref{Vinitialtri}. The calculation of ${\rm Tr}[\hat{V}_{\rm SB}^{0}\hat{\tau}_{\rm SB}]$ follows directly from Eq. \eqref{Vinitialtri} after tracing out the second bath. However, for the calculation of $\lim_{t\rightarrow+\infty}{\rm Tr}[\hat{V}_{\rm SB}^{0}\hat{\Omega}(t_{\rm d},\theta;t)]$, we must also consider that $\ud {\rm Tr}{[\hat{Q}_{0}(t)\hat{P}_{0}(t)\hat{\tau}_{\rm MF}}]/\ud t=0$ holds true, as shown numerically in Appendix \ref{ApB}.

\nocite{*}

\bibliography{apssamp}

\end{document}